\documentclass[aps,prd,twocolumn,superscriptaddress]{revtex4}
\usepackage{epsfig,epsf}
\usepackage{amsmath}
\usepackage{amsthm}
\usepackage{amsfonts}
\usepackage{amssymb}
\usepackage{dsfont}
\usepackage{multirow}
\usepackage{appendix}
\usepackage{slashed}
\usepackage[active]{srcltx}
\usepackage{psfrag}

\begin{document}

\title{Tensor hybrid charmonia }
\date{\today}
\author{S.~S.~Agaev}
\affiliation{Institute for Physical Problems, Baku State University, Az--1148 Baku,
Azerbaijan}
\author{K.~Azizi}
\thanks{Corresponding Author, Email: kazem.azizi@ut.ac.ir}
\affiliation{Department of Physics, University of Tehran, North Karegar Avenue, Tehran
14395-547, Iran}
\affiliation{Department of Physics, Dogus University, Dudullu-\"{U}mraniye, 34775
Istanbul, T\"{u}rkiye}
\author{H.~Sundu}
\affiliation{Department of Physics Engineering, Istanbul Medeniyet University, 34700
Istanbul, T\"{u}rkiye}

\begin{abstract}
The mass and current coupling of the tensor hybrid charmonia $H_{\mathrm{c}}
$ and $\widetilde{H}_{\mathrm{c}}$ with quantum numbers $J^{\mathrm{PC}%
}=2^{-+}$ and $2^{++}$, as well as their full widths are calculated in the
context of the QCD sum rule method. The spectral parameters of these states are
computed using the QCD two-point sum rule approach including dimension-12 terms $%
\sim \langle g_{s}^{3}G^{3}\rangle ^{2} $. The full decay widths of the
charmonia $H_{\mathrm{c}}$ and $\widetilde{H}_{\mathrm{c}}$ are evaluated by
considering their kinematically allowed decay channels. In the case of the
hybrid state $H_{\mathrm{c}} $ decays to $D^{(\pm )}D^{\ast (\mp )}$, $D^{0}%
\overline{D}^{\ast 0}$, and $D_{s}^{(\pm )}D_{s}^{\ast (\mp )} $ mesons are
taken into account. The processes $\widetilde{H}_{\mathrm{c}} \to
D^{(\ast)+}D^{(\ast )-}$, $D^{(\ast) 0}\overline{D}^{(\ast) 0}$, and $%
D_{s}^{(\ast)+}D_{s}^{(\ast) -} $ are employed to estimate the full width of
the hybrid charmonium $\widetilde{H}_{\mathrm{c}}$. The partial widths of
these decays are computed by means of the  QCD three-point sum rule method which
is necessary to calculate strong couplings at the relevant
hybrid-meson-meson vertices. Our predictions $m=(4.16\pm 0.14)~\mathrm{GeV}$%
, and  $\widetilde{m}=(4.5\pm 0.1)~\mathrm{GeV}$ for the masses and $\Gamma \left[
H_{\mathrm{c}}\right] =(160\pm 30)~\mathrm{MeV}$, and  $\Gamma \left[ \widetilde{H%
}_{\mathrm{c}}\right] =(206\pm 33)~\mathrm{MeV}$ for the full width of these
hybrid charmonia can be useful to study and interpret various resonances in
the $4-5~\mathrm{GeV}$ mass range.
\end{abstract}

\maketitle


\section{Introduction}

\label{sec:Intro}

Investigation of structures composed of valence quarks and gluons, i.e.,
hybrid hadrons and their experimental discovery is one of the important problems
in  the agenda of  particle physics. Existence of the hybrids which are
hadrons beyond the conventional $q\overline{q}^{\prime }$ and $qq^{\prime
}q^{\prime \prime }$ scheme is allowed by the quantum chromodynamics and
quark-gluon model. Theoretical studies of such structures have continued over the 
last five decades. Starting from pioneering analyses in Refs.\ \cite%
{Jaffe:1975fd,Horn:1977rq}, the physics of the hybrid mesons and baryons
became a rapidly growing branch of hadronic studies \cite{Meyer:2015eta}.
Numerous publications are devoted to exploring their spectroscopic parameters,
production and decay mechanisms. To this end, researchers suggested new
models and calculational schemes or adapted existing ones to embrace exotic
hadrons as well (see, Refs.\ \cite%
{Meyer:2015eta,Barsbay:2022gtu,Barsbay:2024vjt,Alaakol:2024zyh} and
references therein).

There are few experimentally observed resonances that are considered as
candidates to the hybrid mesons. The light isovector particles $\pi
_{1}(1400)$, $\pi _{1}(1600)$ and $\pi _{1}(2015)$ with the spin-parities $%
J^{\mathrm{PC}}=$ ${1^{-+}}$ are among such structures. It is remarkable
that the ordinary mesons made of a quark and an antidiquark can not bear
such quantum numbers. Therefore, the states $\pi _{1}(1400)$, $\pi
_{1}(1600) $ and $\pi _{1}(2015)$ are definitely exotic particles and
presumably belong to a family of hybrid mesons, though their four-quark
interpretations are not excluded.

The resonance $\pi _{1}(1400)$ has the mass $(1406\pm 20)~\mathrm{MeV}$ and
width $(180\pm 30)~\mathrm{MeV}$. It was seen in the exclusive reaction $\pi
^{-}p\rightarrow \pi ^{0}n\eta $ \cite%
{IHEP-Brussels-LosAlamos-AnnecyLAPP:1988iqi}, and was the first candidate to
the hybrid meson. The next particle from this series, $\pi _{1}(1600)$, was
fixed by $\mathrm{E892}$ collaboration in the decay mode $\eta ^{\prime
}\pi $ of the reaction $\pi ^{-}p$ \cite{E852:2001ikk}. The structures ${%
1^{-+}}$ were observed and studied by this and other experimental groups in
different channels as well \cite%
{CrystalBarrel:1998cfz,E852:2004gpn,E852:2004rfa,COMPASS:2018uzl,COMPASS:2021ogp}%
. The latest analysis, however, favors the existence of only one broad state $%
\pi _{1}(1600)$ \cite{PDG:2024}. The evidence for the next exotic meson $\pi
_{1}(2015)$ from this family was reported in Refs.\ \cite%
{E852:2004gpn,E852:2004rfa}. The isoscalar particle $\eta _{1}(1855)$ with $%
J^{\mathrm{PC}}=$ ${1^{-+}}$ was seen quite recently by  BESIII
collaboration in the radiative decay $J/\psi \rightarrow \gamma \eta
_{1}(1855)\rightarrow \gamma \eta \eta ^{\prime }$ \cite{BESIII:2022riz}.

Some of the observed heavy resonances can be interpreted as candidates to
hybrid mesons as well. For example, it was suggested that the resonances $%
\psi (4230)$ and $\psi (4360)$ may be considered as vector hybrid charmonium
states $\overline{c}gc$ or as mesons with sizeable exotic hybrid ingredients
\cite{Kou:2005gt,Olsen:2017bmm}. A list of numerous other resonances that
probably are hybrid quarkonia was presented in Ref.\ \cite{Brambilla:2022hhi}%
. There are also candidates to hybrid states with baryon quantum numbers. In
fact, the baryon $\Lambda (1405)$ studied by different collaborations \cite%
{Engler:1965zz,Niiyama:2008rt,HADES:2012csk,CLAS:2013rxx,Azizi:2017xyx} may
be one of such exotic baryons (see, for instance, Ref.\ \cite{Azizi:2017xyx}
).

The hybrid quarkonia $H_{\mathrm{b}}=\overline{b}gb$,  and $H_{\mathrm{c}}=%
\overline{c}gc$ and mesons $H_{\mathrm{bc}}=\overline{b}gc$ were
investigated in the frameworks of diverse methods \cite%
{Govaerts:1984hc,Govaerts:1985fx,Page:1996rj,Page:1998gz,Zhu:1998sv,Qiao:2010zh,Harnett:2012gs,HadronSpectrum:2012gic,Chen:2013zia,Chen:2013eha,Berwein:2015vca,Cheung:2016bym,Oncala:2017hop,Palameta:2018yce,Miyamoto:2019oin,Brambilla:2018pyn,Brambilla:2019jfi,Ryan:2020iog,TarrusCastella:2021pld,Woss:2020ayi,Soto:2023lbh,Bruschini:2023tmm,Braaten:2024stn,Wang:2024hvp,TarrusCastella:2024zps,Berwein:2024ztx}%
. These publications addressed numerous problems of heavy hybrid mesons by
studying their spectroscopic parameters, decay channels and production
mechanisms. These states were investigated within the QCD sum rule (SR)
method, the lattice simulations, the nonrelativistic effective field
theories, the Born-Oppenheimer (BO) approximation and BO-effective field
theory (BOEFT).

Results obtained for parameters of hybrid structures in the frameworks of
different methods, as usual, differ from each other. Therefore, there is a
necessity to perform relevant investigations with higher accuracy by
including new factors in the analysis. The hybrid states $H_{\mathrm{b}}$, $H_{%
\mathrm{c}}$, and $H_{\mathrm{bc}}$ were investigated also in our work \cite%
{Alaakol:2024zyh}. There, we calculated the masses and current couplings of
the hybrid quarkonia $H_{\mathrm{b}}$ and $H_{\mathrm{c}}$ with
spin-parities $J^{\mathrm{PC}}=0^{++},\ 0^{+-},\ 0^{-+},\ 0^{--}$ and $%
1^{++},\ 1^{+-},\ 1^{-+},\ 1^{--}$. The spectral parameters of the hybrid
mesons $H_{\mathrm{bc}}$ with $J^{\mathrm{P}}=0^{+}$, $0^{-}$, $1^{+}$, and $%
1^{-}$ were computed as well.

In the current paper we consider the tensor hybrid charmonia $H_{\mathrm{c}}$
and $\widetilde{H}_{\mathrm{c}}$ with contents $\overline{c}gc$ and
spin-parities $J^{\mathrm{PC}}=$ ${2^{-+}}$ and ${2^{++}}$, respectively. We
evaluate their masses and current couplings by employing the QCD two-point
SR method. In this process, we take into account nonperturbative terms $%
\langle g_{s}^{3}G^{3}\rangle ^{2}$. The full decay widths of $H_{\mathrm{c}%
} $ and $\widetilde{H}_{\mathrm{c}}$ are calculated by considering their
kinematically allowed decay channels. It turns out that the hybrid state $H_{%
\mathrm{c}}$ decays to conventional mesons through the processes $H_{\mathrm{%
c}}\rightarrow D^{(\pm )}D^{\ast }(2010)^{(\mp )}$, $D^{0}\overline{D}^{\ast
}(2007)^{0}$, and $D_{s}^{(\pm )}D_{s}^{\ast (\mp )}$. In the case of the
hybrid charmonium $\widetilde{H}_{\mathrm{c}}$ modes $\widetilde{H}_{\mathrm{%
c}}\rightarrow D^{(\ast )+}D^{(\ast )-}$, $D^{(\ast )0}\overline{D}^{(\ast
)0}$, and $D_{s}^{(\ast )+}D_{s}^{(\ast )-}$ are allowed decay channels. The
partial widths of these processes are evaluated by means of QCD three-point
SR approach. This method is necessary to calculate strong couplings at the
relevant hybrid-meson-meson vertices, and in this way, to find the width of
a process under consideration.

This work is divided into five sections. In Sec.\ \ref{sec:Hcharmonium}, we
calculate the masses and current couplings of the tensor mesons $H_{\mathrm{c%
}}$ and $\widetilde{H}_{\mathrm{c}}$. The decay modes of the hybrid state $%
H_{\mathrm{c}}$ are considered in Section \ref{sec:Hcdecays}. The full width
of $\widetilde{H}_{\mathrm{c}}$ is found in Sec.\ \ref{sec:TildeHcWidth}. In
Sec. V we compare our results with those obtained in
other publications and make concluding notes.


\section{Spectroscopic parameters of the tensor hybrids $H_{\mathrm{c}}$ and
$\widetilde{H}_{\mathrm{c}}$}

\label{sec:Hcharmonium}

In this section, we explore the tensor hybrid charmonia $H_{\mathrm{c}}$ and
$\widetilde{H}_{\mathrm{c}}$ by means of the  QCD sum rule method \cite%
{Shifman:1978bx,Shifman:1978by}. The SR method originally was invented to
investigate properties of conventional hadrons, but it can also be employed
to consider multiquark and hybrid particles as well \cite%
{Nielsen:2009uh,Albuquerque:2018jkn,Agaev:2020zad}. It is interesting that
QCD SRs were used for studying the hybrid quarkonia in early years of the
method \cite{Govaerts:1984hc,Govaerts:1985fx}.

We derive the sum rules for the mass $m$ and current coupling $\Lambda $ of
the tensor state $H_{\mathrm{c}}$ using the correlation function
\begin{equation}
\Pi _{\mu \nu \mu ^{\prime }\nu ^{\prime }}(p)=i\int d^{4}xe^{ipx}\langle 0|%
\mathcal{T}\{J_{\mu \nu }(x)J_{\mu ^{\prime }\nu ^{\prime }}^{\dag
}(0)\}|0\rangle ,  \label{eq:CF1}
\end{equation}%
where $J_{\mu \nu }(x)$ and $\mathcal{T}$ \ stand for the interpolating
current of the hybrid state $H_{\mathrm{c}}$ and a time-ordered product of
two currents, respectively.

For the tensor hybrid charmonium with the quantum numbers $J^{\mathrm{PC}%
}=2^{-+}$ the interpolating current is given by the expression
\begin{equation}
J_{\mu \nu }(x)=g_{s}\overline{c}_{a}(x)\sigma _{\mu }^{\alpha }\gamma _{5}%
\frac{{\lambda }_{ab}^{n}}{2}G_{\alpha \nu }^{n}(x)c_{b}(x).  \label{eq:C1}
\end{equation}%
For the tensor state $J^{\mathrm{PC}}=2^{++}$ the interpolating current $%
\widetilde{J}_{\mu \nu }(x)$ has the form
\begin{equation}
\widetilde{J}_{\mu \nu }(x)=g_{s}\overline{c}_{a}(x)\sigma _{\mu }^{\alpha
}\gamma _{5}\frac{{\lambda }_{ab}^{n}}{2}\widetilde{G}_{\alpha \nu
}^{n}(x)c_{b}(x).  \label{eq:C2}
\end{equation}%
In Eqs.\ (\ref{eq:C1}) and (\ref{eq:C2}), $c_{a}(x)$ is the $c$ quark field, and
$g_{s}$ is the QCD strong coupling constant. The $a$ and $b$ are color
indices and ${\lambda }^{n}$, $n=1,2,..8$ stand for the Gell-Mann matrices.
By $G_{\mu \nu }^{n}(x)$ and $\widetilde{G}_{\mu \nu }^{n}(x)=\varepsilon
_{\mu \nu \alpha \beta }G^{n\alpha \beta }(x)/2$ we denote the gluon field
strength tensor and its dual field, respectively.

We start from consideration of the current $J_{\mu \nu }(x)$ and the tensor
hybrid state $H_{\mathrm{c}}$. In the SR method, one first writes $\Pi _{\mu
\nu \mu ^{\prime }\nu ^{\prime }}(p)$ using the physical parameters of $H_{%
\mathrm{c}}$
\begin{eqnarray}
&&\Pi _{\mu \nu \mu ^{\prime }\nu ^{\prime }}^{\mathrm{Phys}}(p)=\frac{%
\langle 0|J_{\mu \nu }|H_{\mathrm{c}}(p,\epsilon )\rangle \langle H_{\mathrm{%
c}}(p,\epsilon )|J_{\mu ^{\prime }\nu ^{\prime }}^{\dag }|0\rangle }{%
m^{2}-p^{2}}  \notag \\
&&+\cdots ,  \label{eq:CF2}
\end{eqnarray}%
where $m$ and $\epsilon =\epsilon _{\mu \nu }^{(\lambda )}(p)$ are its mass
and polarization tensor, respectively. Here, the contribution of the
ground-level particle $H_{\mathrm{c}}$ is shown explicitly, whereas effects
due to higher resonances and continuum states are denoted by the ellipses.
It is also useful to introduce the matrix element
\begin{equation}
\langle 0|J_{\mu \nu }|H_{\mathrm{c}}(p,\epsilon )\rangle =\Lambda \epsilon
_{\mu \nu }^{(\lambda )}(p).  \label{eq:M1}
\end{equation}%
To find $\Pi _{\mu \nu \mu ^{\prime }\nu ^{\prime }}^{\mathrm{Phys}}(p)$, we
substitute this matrix element into Eq.\ (\ref{eq:CF2}) and perform
summation over the polarization tensor using
\begin{eqnarray}
\sum\limits_{\lambda }\epsilon _{\mu \nu }^{(\lambda )}(p)\epsilon _{\mu
^{\prime }\nu ^{\prime }}^{\ast (\lambda )}(p) &=&\frac{1}{2}(\widetilde{g}%
_{\mu \mu ^{\prime }}\widetilde{g}_{\nu \nu ^{\prime }}+\widetilde{g}_{\mu
\nu ^{\prime }}\widetilde{g}_{\nu \mu ^{\prime }})  \notag \\
&&-\frac{1}{3}\widetilde{g}_{\mu \nu }\widetilde{g}_{\mu ^{\prime }\nu
^{\prime }},  \label{eq:F1}
\end{eqnarray}%
where%
\begin{equation}
\widetilde{g}_{\mu \nu }=-g_{\mu \nu }+\frac{p_{\mu }p_{\nu }}{p^{2}}.
\label{eq:F2}
\end{equation}%
Our computations yield
\begin{eqnarray}
\Pi _{\mu \nu \mu ^{\prime }\nu ^{\prime }}^{\mathrm{Phys}}(p) &=&\frac{%
\Lambda ^{2}}{m^{2}-p^{2}}\left\{ \frac{1}{2}\left( g_{\mu \mu ^{\prime
}}g_{\nu \nu ^{\prime }}+g_{\mu \nu ^{\prime }}g_{\nu \mu ^{\prime }}\right)
\right.  \notag \\
&&\left. +\ \text{other structures}\right\} +..,  \label{eq:Phys2}
\end{eqnarray}%
with dots standing for contributions of other structures as well as higher
resonances and continuum states. Note that, after application of Eqs.\ (\ref%
{eq:F1}) and (\ref{eq:F2}) there appear numerous Lorentz structures in the
curly brackets. The term proportional to $(g_{\mu \mu ^{\prime }}g_{\nu \nu
^{\prime }}+g_{\mu \nu ^{\prime }}g_{\nu \mu ^{\prime }})$ contains
contribution of only the spin-$2$ particle, whereas the remaining components in Eq.\
(\ref{eq:Phys2}) are formed due to contributions of spin-$0$ and -$1$ states
as well. Therefore, in our studies we restrict ourselves by exploring this
term and the corresponding invariant amplitude $\Pi ^{\mathrm{Phys}}(p^{2})$.

At the next stage of analysis, we compute the correlator $\Pi _{\mu \nu \mu
^{\prime }\nu ^{\prime }}(p)$ with some accuracy in the operator product
expansion ($\mathrm{OPE}$). To this end, we employ in Eq.\ (\ref{eq:CF1})
expression of the current $J_{\mu \nu }(x)$ and contract corresponding quark
and gluon fields. As a result, we find
\begin{eqnarray}
&&\Pi _{\mu \nu \mu ^{\prime }\nu ^{\prime }}^{\mathrm{OPE}}(p)=\frac{%
ig_{s}^{2}}{4}\int d^{4}xe^{ipx}{\lambda }_{ab}^{n}{\lambda }_{a^{\prime
}b^{\prime }}^{m}\langle 0|G_{\alpha \nu }^{n}(x)G_{\alpha ^{\prime }\nu
^{\prime }}^{m}(0)|0\rangle  \notag \\
&&\times \mathrm{Tr}\left[ \sigma _{\mu }^{\alpha }\gamma
_{5}S_{c}^{bb^{\prime }}(x)\gamma _{5}\sigma _{\mu ^{\prime }}^{\alpha
^{\prime }}S_{c}^{a^{\prime }a}(-x)\right] ,  \label{eq:OPE1}
\end{eqnarray}%
where $S_{c}^{ab}(x)$ is the $c$ quark propagator. In our calculations we
employ the following expression for $S_{c}^{ab}(x)$ [$Q=c$]
\begin{eqnarray}
&&S_{Q}^{ab}(x)=i\int \frac{d^{4}k}{(2\pi )^{4}}e^{-ikx}\Bigg \{\frac{\delta
_{ab}\left( {\slashed k}+m_{Q}\right) }{k^{2}-m_{Q}^{2}}  \notag \\
&&-\frac{g_{s}G_{ab}^{\alpha \beta }}{4}\frac{\sigma _{\alpha \beta }\left( {%
\slashed k}+m_{Q}\right) +\left( {\slashed k}+m_{Q}\right) \sigma _{\alpha
\beta }}{(k^{2}-m_{Q}^{2})^{2}}  \notag \\
&&+\frac{g_{s}^{2}G^{2}}{12}\delta _{ab}m_{Q}\frac{k^{2}+m_{Q}{\slashed k}}{%
(k^{2}-m_{Q}^{2})^{4}}+\frac{g_{s}^{3}G^{3}}{48}\delta _{ab}\frac{\left( {%
\slashed k}+m_{Q}\right) }{(k^{2}-m_{Q}^{2})^{6}}  \notag \\
&&\times \left[ {\slashed k}\left( k^{2}-3m_{Q}^{2}\right) +2m_{Q}\left(
2k^{2}-m_{Q}^{2}\right) \right] \left( {\slashed k}+m_{Q}\right) +\cdots %
\Bigg \}.  \notag \\
&&  \label{eq:Bprop}
\end{eqnarray}%
Here, we have used the short-hand notations
\begin{eqnarray}
G_{ab}^{\alpha \beta } &\equiv &G_{n}^{\alpha \beta }\lambda _{ab}^{n}/2,\ \
G^{2}=G_{\alpha \beta }^{n}G_{n}^{\alpha \beta },\   \notag \\
G^{3} &=&f^{nml}G_{\alpha \beta }^{n}G^{m\beta \delta }G_{\delta }^{l\alpha
},
\end{eqnarray}%
where $f^{nml}$ are the structure constants of the color group $SU_{c}(3)$.

The $\Pi _{\mu \nu \mu ^{\prime }\nu ^{\prime }}^{\mathrm{OPE}}(p)$ is a
product of two factors. One of them is the term with the trace over spinor
indices and consists of $c$ quark propagators. The propagator $S_{c}^{ab}(x)$
has the perturbative and nonperturbative components. The latter includes
terms proportional to $g_{s}^{2}G^{2}$ and $g_{s}^{3}G^{3}$ which between
vacuum states generate the well known gluon condensates. The term $\sim
g_{s}G_{ab}^{\alpha \beta }$ in Eq.\ (\ref{eq:Bprop}), having been  multiplied with
a similar component of another propagator gives rise to the two-gluon
condensate as well, which we also include in the analysis.

Our treatment of the matrix element $\langle 0|G_{\alpha \nu
}^{n}(x)G_{\alpha ^{\prime }\nu ^{\prime }}^{m}(0)|0\rangle $ also needs
some comments. We replace it with the vacuum condensate $\langle
g_{s}^{2}G^{2}\rangle $ and keep the first component in the Taylor expansion
at $x=0$
\begin{eqnarray}
&&\langle 0|g_{s}^{2}G_{\alpha \nu }^{n}(x)G_{\alpha ^{\prime }\nu ^{\prime
}}^{m}(0)|0\rangle =\frac{\langle g_{s}^{2}G^{2}\rangle }{96}\delta ^{nm}
\notag \\
&&\times \left[ g_{\alpha \alpha ^{\prime }}g_{\nu \nu ^{\prime }}-g_{\alpha
\nu ^{\prime }}g_{\alpha ^{\prime }\nu }\right] .  \label{eq:GluonME1}
\end{eqnarray}%
Terms obtained in this manner describe diagrams where the gluon interacts
with the QCD vacuum. Alternatively, we use
\begin{eqnarray}
&&\langle 0|G_{\alpha \nu }^{n}(x)G_{\alpha ^{\prime }\nu ^{\prime
}}^{m}(0)|0\rangle =\frac{\delta ^{nm}}{2\pi ^{2}x^{4}}\left[ g_{\nu \nu
^{\prime }}\left( g_{\alpha \alpha ^{\prime }}-\frac{4x_{\alpha }x_{\alpha
^{\prime }}}{x^{2}}\right) \right.  \notag \\
&&\left. +(\nu ,\nu ^{\prime })\leftrightarrow (\alpha ,\alpha ^{\prime
})-\nu \leftrightarrow \alpha -\nu ^{\prime }\leftrightarrow \alpha ^{\prime
}\right] .  \label{eq:GluonME2}
\end{eqnarray}%
Contributions obtained in this way correspond to diagrams with the full valence
gluon propagator.

Having extracted the structure $(g_{\mu \mu ^{\prime }}g_{\nu \nu ^{\prime
}}+g_{\mu \nu ^{\prime }}g_{\nu \mu ^{\prime }})$ from $\Pi _{\mu \nu \mu
^{\prime }\nu ^{\prime }}^{\mathrm{OPE}}(p)$ and labeled the corresponding
invariant amplitude by $\Pi ^{\mathrm{OPE}}(p^{2})$, one derives SRs for the
mass and current coupling of the hybrid meson $H_{\mathrm{c}}$. To this end,
we rewrite $\Pi ^{\mathrm{Phys}}(p^{2})$ as the dispersion integral%
\begin{equation}
\Pi ^{\mathrm{Phys}}(p^{2})=\int_{4m_{c}^{2}}^{\infty }\frac{\rho ^{\mathrm{%
Phys}}(s)ds}{s-p^{2}}+\cdots ,  \label{eq:DisRel}
\end{equation}%
where $m_{c}$ is the $c$ quark mass, and the dots indicate subtraction terms
required to render finite $\Pi ^{\mathrm{Phys}}(p^{2})$. The spectral
density $\rho ^{\mathrm{Phys}}(s)$ is equal to the imaginary part of $\Pi ^{%
\mathrm{Phys}}(p^{2})$,%
\begin{equation}
\rho ^{\mathrm{Phys}}(s)=\Lambda ^{2}\delta (s-m^{2})+\rho ^{\mathrm{h}%
}(s)\theta (s-s_0),  \label{eq:SDensity}
\end{equation}%
where $s_{0}$ is the continuum subtraction parameter. The contribution of
the hybrid meson $H_{\mathrm{c}}$ is represented in Eq.\ (\ref{eq:SDensity})
by the pole term and separated from other effects. Contributions to $\rho ^{%
\mathrm{Phys}}(s)$ coming from higher resonances and continuum states are
characterized by an unknown hadronic spectral density $\rho ^{\mathrm{h}}(s)$%
. It is clear that $\rho ^{\mathrm{Phys}}(s)$ leads to the expression
\begin{equation}
\Pi ^{\mathrm{Phys}}(p^{2})=\frac{\Lambda ^{2}}{m^{2}-p^{2}}%
+\int_{s_0}^{\infty }\frac{\rho ^{\mathrm{h}}(s)ds}{s-p^{2}}+\cdots .
\label{eq:InvAmp2}
\end{equation}%
We employ, in the region $p^{2}\ll 0$, the Borel transformation $\mathcal{B}$
to suppress contributions of higher resonances and continuum states. This
transformation also removes subtraction terms in the dispersion integral.
For $\mathcal{B}\Pi ^{\mathrm{Phys}}(p^{2})$, we obtain%
\begin{equation}
\mathcal{B}\Pi ^{\mathrm{Phys}}(p^{2})=\Lambda
^{2}e^{-m^{2}/M^{2}}+\int_{s_0}^{\infty }ds\rho ^{\mathrm{h}}(s)e^{-s/M^{2}}.
\label{eq:CorBor}
\end{equation}%
where $M^{2}$ is the Borel parameter.

The amplitude $\Pi ^{\mathrm{OPE}}(p^{2})$ can be calculated in the deep
Euclidean region $p^{2}\ll 0$ using the operator product expansion. The
coefficient functions in $\mathrm{OPE}$ could be obtained using methods of
perturbative QCD, whereas nonperturbative information is contained in the
gluon condensates. Having calculated the imaginary part of $\Pi ^{\mathrm{OPE%
}}(p^{2})$, we get the two-point spectral density $\rho ^{\mathrm{OPE}}(s)$.

One can also write the dispersion representation for the amplitude $\Pi ^{%
\mathrm{OPE}}(p^{2})$ in terms of $\rho ^{\mathrm{OPE}}(s)$. Then, by
equating the Borel transformations of $\Pi ^{\mathrm{Phys}}(p^{2})$ and $\Pi
^{\mathrm{OPE}}(p^{2})$ and using the assumption on hadron-parton duality $%
\rho ^{\mathrm{h}}(s)\simeq \rho ^{\mathrm{OPE}}(s)$ above the threshold, we
can remove the second term in Eq.\ (\ref{eq:CorBor}) from the right-hand
side of the obtained equality and find
\begin{equation}
\Lambda ^{2}e^{-m^{2}/M^{2}}=\Pi (M^{2},s_{0}).  \label{eq:SR}
\end{equation}%
Here,
\begin{equation}
\Pi (M^{2},s_{0})=\int_{4m_{c}^{2}}^{s_{0}}ds\rho ^{\mathrm{OPE}%
}(s)e^{-s/M^{2}}.  \label{eq:CorrF}
\end{equation}

After simple manipulations, we get
\begin{equation}
m^{2}=\frac{\Pi ^{\prime }(M^{2},s_{0})}{\Pi (M^{2},s_{0})},  \label{eq:Mass}
\end{equation}%
and
\begin{equation}
\Lambda ^{2}=e^{m^{2}/M^{2}}\Pi (M^{2},s_{0}),  \label{eq:Coupl}
\end{equation}%
which are the sum rules for $m$ and $\Lambda $, respectively. In Eq.\ (\ref%
{eq:Mass}), we have introduced the notation $\Pi ^{\prime
}(M^{2},s_{0})=d\Pi (M^{2},s_{0})/d(-1/M^{2})$. The spectral density $\rho ^{%
\mathrm{OPE}}(s)$ contains the perturbative $\rho ^{\mathrm{pert.}}(s)$ and
nonperturbative $\rho ^{\mathrm{DimN}}(s)$ terms ($\mathrm{N=4,6,8,10,12}$).
Explicit expressions for $\rho ^{\mathrm{OPE}}(s)$ are lengthy and not
provided here.

To carry out numerical analysis, we have to fix the input parameters in
Eqs.\ (\ref{eq:Mass}) and (\ref{eq:Coupl}). For these purposes, we employ
the values
\begin{eqnarray}
&&m_{c}=(1.27\pm 0.02)~\mathrm{GeV},  \notag \\
&&\langle \alpha _{s}G^{2}/\pi \rangle =(0.012\pm 0.004)~\mathrm{GeV}^{4},
\notag \\
&&\langle g_{s}^{3}G^{3}\rangle =(0.57\pm 0.29)~\mathrm{GeV}^{6}.
\label{eq:GluonCond}
\end{eqnarray}%
Here, $m_{c}$ corresponds to the running mass in the $\overline{\mathrm{MS}}$
scheme at the scale $\mu =m_{c}$ \cite{PDG:2024}. The condensates $\langle
\alpha _{s}G^{2}/\pi \rangle $ and $\langle g_{s}^{3}G^{3}\rangle $ were
obtained from analyses of different processes \cite%
{Shifman:1978bx,Shifman:1978by,Narison:2015nxh}.

Equations (\ref{eq:Mass}) and (\ref{eq:Coupl}) also contain  the parameters $%
M^{2}$ and $s_{0}$. They have to be chosen in such a way as to guarantee
the prevalence of the pole contribution ($\mathrm{PC}$) in the physical
quantities obtained from the SR method. The stability of these results on $%
M^{2}$, as well as convergence of the $\mathrm{OPE}$ are also important
restrictions in numerical calculations. Numerically, these conditions can be
controlled by introducing the quantities
\begin{equation}
\mathrm{PC}=\frac{\Pi (M^{2},s_{0})}{\Pi (M^{2},\infty )},  \label{eq:PC}
\end{equation}%
and%
\begin{equation}
R(M^{2})=\frac{\Pi ^{\mathrm{DimN}}(M^{2},s_{0})}{\Pi (M^{2},s_{0})},
\label{eq:Conv}
\end{equation}%
where $\Pi ^{\mathrm{DimN}}(M^{2},s_{0})=\sum_{\mathrm{N}=8,10,12}\Pi ^{%
\mathrm{DimN}}$ is a sum of terms proportional to $\langle
g_{s}^{2}G^{2}\rangle ^{2}$,\ $\langle g_{s}^{2}G^{2}\rangle \langle
g_{s}^{3}G^{3}\rangle $ and $\langle g_{s}^{3}G^{3}\rangle ^{2}$,
respectively.

\begin{figure}[h]
\includegraphics[width=8.5cm]{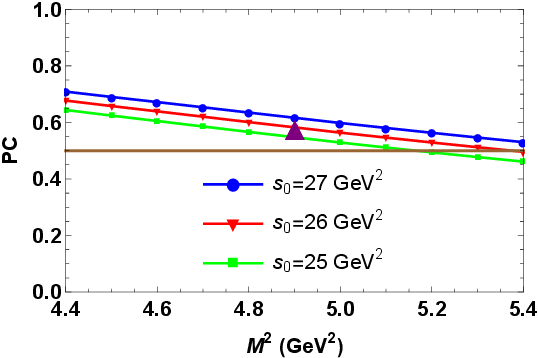}
\caption{Pole contribution $\mathrm{PC}$ as a function of $M^{2}$ at fixed $%
s_{0}$. The red triangle marks the point $M^{2}=4.9~\mathrm{GeV}^{2}$ and $%
s_{0}=26~\mathrm{GeV}^{2}$. }
\label{fig:PC}
\end{figure}

Our numerical computations prove that in the case of $H_{\mathrm{c}}$ the
regions
\begin{equation}
M^{2}\in \lbrack 4.4,5.4]~\mathrm{GeV}^{2},\ s_{0}\in \lbrack 25,27]~\mathrm{%
GeV}^{2},  \label{eq:Wind1}
\end{equation}%
satisfy restrictions of the SR analysis. Indeed, at $M^{2}=4.4~\mathrm{GeV}%
^{2}$ and $M^{2}=5.4~\mathrm{GeV}^{2}$ the averaged value of the pole
contribution is $\mathrm{PC}\approx 0.68$ and $\mathrm{PC}$ $\approx 0.50$,
respectively. At $M^{2}=4.4~\mathrm{GeV}^{2}$ the parameter $\left\vert
R(M^{2})\right\vert $ does not exceed $0.01$. The  $\mathrm{%
PC}$ is shown in Fig.\ \ref{fig:PC} as a function of $M^{2}$.

The spectral parameters $m$ and $\Lambda $ are found as their average values
over the regions in  Eq.\ (\ref{eq:Wind1}) and amount to
\begin{eqnarray}
m &=&(4.16\pm 0.14)~\mathrm{GeV},\   \notag \\
\Lambda &=&(0.68\pm 0.04)~\mathrm{GeV}^{4}.  \label{eq:Result1}
\end{eqnarray}%
The predictions in Eq.\ (\ref{eq:Result1}) correspond to the SR results at
the point $M^{2}=4.9~\mathrm{GeV}^{2}$ and $s_{0}=26~\mathrm{GeV}^{2}$,
where the pole contribution is $\mathrm{PC}\approx 0.58$. This fact ensures
the dominance of the pole contribution, and demonstrates the ground-state nature
of $H_{\mathrm{c}}$.

\begin{widetext}

\begin{figure}[h]
\begin{center}
\includegraphics[totalheight=6cm,width=8cm]{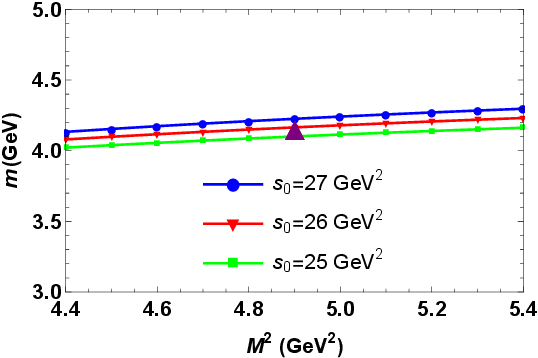} %
\includegraphics[totalheight=6cm,width=8cm]{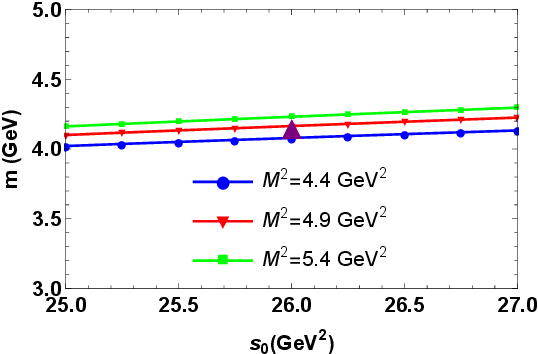}
\end{center}
\caption{Mass $m$ vs $M^{2}$ (left) and $m$ vs $s_{0}$
(right). The triangles on the plots show the position of $H_{\mathrm{c}}$.}
\label{fig:Mass1}
\end{figure}

\end{widetext}

Errors in Eq.\ (\ref{eq:Result1}) are generated mainly by the parameters $%
M^{2}$ and $s_{0}$ and the gluon condensate $\langle \alpha _{s}G^{2}/\pi
\rangle $. Uncertainties of the condensate $\langle g_{s}^{3}G^{3}\rangle $
lead to corrections, which in the case of $m$ are very small. Throughout
this work, in calculations we use the central value   for  $\langle g_{s}^{3}G^{3}\rangle $. The mass $m$ is shown in Fig.\ \ref{fig:Mass1} as a function
of the parameters $M^{2}$ and $s_{0}$.

The hybrid tensor meson $\widetilde{H}_{\mathrm{c}}$ with the spin-parities $%
J^{\mathrm{PC}}=2^{++}$ is explored in the same manner. In this case $%
\widetilde{\Pi }_{\mu \nu \mu ^{\prime }\nu ^{\prime }}^{\mathrm{OPE}}(p)$
is obtained from Eq.\ (\ref{eq:OPE1}) after substitution $G\rightarrow
\widetilde{G}$. The working regions for the parameters $M^{2}$ and $s_{0}$
are
\begin{equation}
M^{2}\in \lbrack 4,5]~\mathrm{GeV}^{2},\ s_{0}\in \lbrack 26,28]~\mathrm{GeV}%
^{2}.  \label{eq:Wind1A}
\end{equation}%
In these regions the pole contribution changes inside of the limits $%
0.50\leq \mathrm{PC}\leq 0.68$. As a result, for the mass $\widetilde{m}$
and current coupling $\widetilde{\Lambda }$ of the hybrid state $\widetilde{H%
}_{\mathrm{c}}$, we get
\begin{eqnarray}
\widetilde{m} &=&(4.5\pm 0.1)~\mathrm{GeV},\ \   \notag \\
\widetilde{\Lambda } &=&(0.27\pm 0.04)~\mathrm{GeV}^{4}.
\end{eqnarray}%
These results correspond to SR predictions at the point $M^{2}=4.5~\mathrm{%
GeV}^{2}$ and $s_{0}=27~\mathrm{GeV}^{2}$, where $\mathrm{PC}\approx 0.59$.


\section{Decays of the hybrid meson $H_{\mathrm{c}}$}

\label{sec:Hcdecays}

In this section we consider decays of the hybrid charmonium $H_{\mathrm{c}}$%
. There are two-body processes $H_{\mathrm{c}}\rightarrow D^{(\pm )}D^{\ast
}(2010)^{(\mp )}$, $D^{0}\overline{D}^{\ast }(2007)^{0}$, and $D_{s}^{(\pm
)}D_{s}^{\ast (\mp )}$ which are kinematically allowed modes of $H_{\mathrm{c%
}}$. In fact, it is not difficult to see that the mass $m$ of $H_{\mathrm{c}%
} $ exceeds the two-meson thresholds for these decays. The partial widths of
these channels apart from other factors are determined by the strong
couplings $g_{i}$ at the corresponding $H_{\mathrm{c}}$-meson-meson
vertices. To evaluate $g_{i}$ we employ the three-point SR method enabling
us to extract information on the strong form factors $g_{i}(q^{2})$ which at
the relevant mass shells $q^{2}=m_{D}^{2}$ become equal to the couplings of
interest.


\subsection{$H_{\mathrm{c}}\rightarrow D^{+}D^{\ast }(2010)^{-}$, $%
D^{-}D^{\ast }(2010)^{+}$ and $D^{0}\overline{D}^{\ast }(2007)^{0}$}


Here, we are going to analyze in expanded form the process $H_{\mathrm{c}%
}\rightarrow D^{+}D^{\ast }(2010)^{-}$ and calculate its partial width $%
\Gamma \lbrack H_{\mathrm{c}}\rightarrow D^{+}D^{\ast }{}^{-}]$. The partial
width of the second decay is equal to $\Gamma \lbrack H_{\mathrm{c}%
}\rightarrow D^{+}D^{\ast }{}^{-}]$ as well. The quark content $\overline{u}%
c+(\overline{c}u)^{\ast }$ of the mode $D^{0}\overline{D}^{\ast }(2007)^{0}$
after replacement $u\rightarrow d$ becomes $\overline{d}c+(\overline{c}%
d)^{\ast }\rightarrow D^{+}D^{\ast }{}^{-}$. In the present work we adopt an
approximation $m_{u}=m_{d}=0$, and use the same decay constants for the
charged and neutral $D^{(\ast )}$ particles. The differences between the
masses of the $D^{0}$, $\overline{D}^{\ast }(2007)^{0}$ and $D^{+}$, $%
D^{\ast }(2010)^{-}$ mesons are small and neglected in this work. Therefore,
with high accuracy the partial decay width of the process $H_{\mathrm{c}%
}\rightarrow D^{0}\overline{D}^{\ast }(2007)^{0}$ is equal to that of the
first decay.

The coupling $g_{1}$ that describes strong interaction of particles at the
vertex $H_{\mathrm{c}}D^{+}D^{\ast -}$ can be evaluated using the
three-point correlation function. First, we find the sum rule for the form
factor $g_{1}(q^{2})$ and consider, to this end, the correlation function
\begin{eqnarray}
\Pi _{\mu \alpha \beta }(p,p^{\prime }) &=&i^{2}\int
d^{4}xd^{4}ye^{ip^{\prime }x}e^{iqy}\langle 0|\mathcal{T}\{J_{\mu }^{D^{\ast
}}(x)  \notag \\
&&\times J^{D}(y)J_{\alpha \beta }^{\dagger }(0)\}|0\rangle .
\label{eq:CF1a}
\end{eqnarray}%
Here, $J_{\mu }^{D^{\ast }}(x)$ and $J^{D}(x)$ are interpolating currents of
the vector $D^{\ast }(2010)^{-}$ and pseudoscalar $D^{+}$ mesons,
respectively. They are defined as
\begin{equation}
J_{\mu }^{D^{\ast }}(x)=\overline{c}_{i}(x)\gamma _{\mu }d_{i}(x),\ J^{D}(x)=%
\overline{d}_{j}(x)i\gamma _{5}c_{j}(x),  \label{eq:CDmes}
\end{equation}%
with $i$ and $j$ being the color indices.

To determine the phenomenological side $\Pi _{\mu \alpha \beta }^{\mathrm{%
Phys}}(p,p^{\prime })$ of the sum rule, we rewrite Eq.\ (\ref{eq:CF1a}) in
terms of the particles' parameters. By taking into account only
contributions of the ground-state particles, we recast the correlator $\Pi
_{\mu \alpha \beta }(p,p^{\prime })$ into the form
\begin{eqnarray}
&&\Pi _{\mu \alpha \beta }^{\mathrm{Phys}}(p,p^{\prime })=\frac{\langle
0|J_{\mu }^{D^{\ast }}|D^{\ast -}(p^{\prime },\varepsilon )\rangle }{%
p^{\prime 2}-m_{D^{\ast }}^{2}}\frac{\langle 0|J^{D}|D^{+}(q)\rangle }{%
q^{2}-m_{D}^{2}}  \notag \\
&&\times \langle D^{\ast -}(p^{\prime },\varepsilon )D^{+}(q)|H_{\mathrm{c}%
}(p,\epsilon )\rangle \frac{\langle H_{\mathrm{c}}(p,\epsilon )|J_{\alpha
\beta }^{\dagger }|0\rangle }{p^{2}-m^{2}}+\cdots ,  \notag \\
&&  \label{eq:TP1}
\end{eqnarray}%
where $m_{D^{\ast }}=(2010.26\pm 0.05)~\mathrm{MeV}$ and $m_{D}=(1869.5\pm
0.4)~\mathrm{MeV}$ are the masses of the mesons $D^{\ast -}$ and $D^{+}$
\cite{PDG:2024}, respectively. Above, we denote by $\varepsilon _{\mu
}(p^{\prime })$ the polarization vector of the meson $D^{\ast }{}^{-}$.

To further simplify Eq.\ (\ref{eq:TP1}), we introduce the following matrix
elements
\begin{eqnarray}
\langle 0|J_{\mu }^{D^{\ast }}|D^{\ast -}(p^{\prime },\varepsilon )\rangle
&=&f_{D^{\ast }}m_{D^{\ast }}\varepsilon _{\mu }(p^{\prime }),  \notag \\
\langle 0|J^{D}|D^{+}(q)\rangle &=&\frac{f_{D}m_{D}^{2}}{m_{c}},
\label{eq:ME1}
\end{eqnarray}%
where $f_{D^{\ast }}=(223.5\pm 8.4)~\mathrm{MeV}$ and $f_{D}=(212.0\pm 0.7)~%
\mathrm{MeV}$ are decay constants of the mesons. We have  to also fix the
matrix element $\langle D^{\ast -}(p^{\prime },\varepsilon )D(q)|H_{\mathrm{c%
}}(p,\epsilon )\rangle $ that is given by the expression
\begin{eqnarray}
\langle D^{\ast -}(p^{\prime },\varepsilon )D(q)|H_{\mathrm{c}}(p,\epsilon
)\rangle &=&g_{1}(q^{2})\epsilon ^{\rho \sigma }(p)q_{\rho }\varepsilon
_{\sigma }(p^{\prime }).  \notag \\
&&
\end{eqnarray}

As a result, for $\Pi _{\mu \alpha \beta }^{\mathrm{Phys}}(p,p^{\prime })$
we get
\begin{eqnarray}
&&\Pi _{\mu \alpha \beta }^{\mathrm{Phys}}(p,p^{\prime })=g_{1}(q^{2})\frac{%
\Lambda f_{D^{\ast }}m_{D^{\ast }}f_{D}m_{D}^{2}}{m_{c}\left(
p^{2}-m^{2}\right) (p^{\prime 2}-m_{D^{\ast }}^{2})}  \notag \\
&&\times \frac{1}{(q^{2}-m_{D}^{2})}\left[ \frac{1}{2}g_{\mu \alpha
}p_{\beta }^{\prime }-\frac{(m^{2}+m_{D^{\ast }}^{2}-q^{2})^{2}}{%
12m^{2}m_{D^{\ast }}^{2}}g_{\alpha \beta }p_{\mu }^{\prime }\right.  \notag
\\
&&\left. +\ \text{other structures}\right] +\cdots .  \label{eq:PhysSide}
\end{eqnarray}

For $\Pi _{\mu \alpha \beta }^{\mathrm{OPE}}(p,p^{\prime })$ one finds%
\begin{eqnarray}
&&\Pi _{\mu \alpha \beta }^{\mathrm{OPE}}(p,p^{\prime })=-i\int
d^{4}xd^{4}ye^{ip^{\prime }x}e^{iqy}g_{s}\frac{{\lambda }_{ab}^{n}}{2}%
G_{\rho \alpha }^{n}(0)  \notag \\
&&\times \mathrm{Tr}\left[ \gamma _{\beta }S_{d}^{ij}(x-y)\gamma
_{5}S_{c}^{jb}(y)\gamma _{5}\sigma _{\mu }^{\rho }S_{c}^{ai}(-x)\right] ,
\label{eq:CF3}
\end{eqnarray}%
where $S_{d}^{ij}(x-y)$ is the light quark propagator \cite{Agaev:2020zad}.
To find SR for the form factor $g_{1}(q^{2})$, we utilize the structures
proportional to $g_{\mu \alpha }p_{\beta }^{\prime }$ in the correlation
functions and corresponding amplitudes $\Pi _{1}^{\mathrm{Phys}%
}(p^{2},p^{\prime 2})$ and $\Pi _{1}^{\mathrm{OPE}}(p^{2},p^{\prime 2})$.
After standard operations the sum rule for $g_{1}(q^{2})$ reads
\begin{equation}
g_{1}(q^{2})=\frac{2m_{c}(q^{2}-m_{D}^{2})}{\Lambda f_{D^{\ast }}m_{D^{\ast
}}f_{D}m_{D}^{2}}e^{m^{2}/M_{1}^{2}}e^{m_{D^{\ast }}^{2}/M_{2}^{2}}\Pi _{1}(%
\mathbf{M}^{2},\mathbf{s}_{0}).  \label{eq:SRG}
\end{equation}%
In Eq.\ (\ref{eq:SRG}) $\Pi _{1}(\mathbf{M}^{2},\mathbf{s}_{0})$ is the
Borel transformed and subtracted function $\Pi _{1}^{\mathrm{OPE}%
}(p^{2},p^{\prime 2})$. It depends on the parameters $\mathbf{M}%
^{2}=(M_{1}^{2},M_{2}^{2})$ and $\mathbf{s}_{0}=(s_{0},s_{0}^{\prime })$
where the pairs $(M_{1}^{2},s_{0})$ and $(M_{2}^{2},s_{0}^{\prime })$
correspond to the hybrid $H_{\mathrm{c}}$ and $D^{\ast }{}^{-}$ channels,
respectively. The function $\Pi _{1}(\mathbf{M}^{2},\mathbf{s}_{0})$ amounts
to
\begin{eqnarray}
&&\Pi _{1}(\mathbf{M}^{2},\mathbf{s}_{0})=\int_{4m_{c}^{2}}^{s_{0}}ds%
\int_{m_{c}^{2}}^{s_{0}^{\prime }}ds^{\prime }\rho _{1}(s,s^{\prime })
\notag \\
&&\times e^{-s/M_{1}^{2}-s^{\prime }/M_{2}^{2}}.  \label{eq:CorrF1}
\end{eqnarray}%
The spectral density $\rho _{1}(s,s^{\prime })$ is evaluated as the
imaginary part of the amplitude $\Pi _{1}^{\mathrm{OPE}}(p^{2},p^{\prime 2})$%
. Calculations are performed by taking into account nonperturbative terms up
to dimension $10$. \ The spectral density $\rho _{1}(s,s^{\prime })$ has the
components $\rho _{1}^{\mathrm{pert.}}(s,s^{\prime })$ and $\rho _{1}^{%
\mathrm{Dim4}}(s,s^{\prime })$ which are given by the formulas,
\begin{equation}
\rho _{1}^{\mathrm{pert.}}(s,s^{\prime })=\frac{g_{s}^{2}m_{c}^{2}}{32\pi
^{2}}\int_{0}^{1}d\alpha \int_{0}^{1-\alpha }\frac{d\beta \theta (N)N^{2}}{%
\alpha \beta ^{3}(\alpha +\beta -1)},  \label{eq:RhoPert}
\end{equation}%
and%
\begin{eqnarray}
&&\rho _{1}^{\mathrm{Dim4}}(s,s^{\prime })=\frac{\langle \alpha
_{s}G^{2}/\pi \rangle m_{c}^{2}}{12}\int_{0}^{1}d\alpha \int_{0}^{1-\alpha }%
\frac{d\beta \theta (N)}{\beta ^{5}}  \notag \\
&&\times \left[ g_{s}^{2}\frac{\alpha (\alpha ^{4}+\beta ^{4})}{12\beta
^{2}(\alpha +\beta -1)}+\pi ^{2}\alpha (1-\alpha )\right] ,
\label{eq:RhoDim4}
\end{eqnarray}%
In Eqs.\ (\ref{eq:RhoPert}) and (\ref{eq:RhoDim4}) $\theta (N)$ is the unit
step function with the argument
\begin{equation}
N=\frac{s^{\prime }-(\alpha +\beta )(s^{\prime }+m_{c}^{2})}{\beta }.
\end{equation}%
As is seen, only the perturbative and dimension-$4$ terms contribute to the
correlation function $\Pi _{1}(\mathbf{M}^{2},\mathbf{s}_{0})$. The limited
number of the terms in $\rho _{1}(s,s^{\prime })$ is connected with the
forms of the interpolating currents for $H_{\mathrm{c}}$, $D^{+}$ and $%
D^{\ast }{}^{-}$, mathematical manipulations used to derive $\rho
_{1}(s,s^{\prime })$, and with the approximation $m_{u}=m_{d}=0$ adopted in
the present work.

In numerical computations, we choose the parameters $(M_{1}^{2},s_{0})$ and $%
(M_{2}^{2},s_{0}^{\prime })$ in the following manner: In the hybrid channels
we use $(M_{1}^{2},s_{0})$ from Eq.\ (\ref{eq:Wind1}), whereas for the $%
D^{\ast }{}^{-}$ meson channel, we employ
\begin{equation}
M_{2}^{2}\in \lbrack 2,4]~\mathrm{GeV}^{2},\ s_{0}^{\prime }\in \lbrack
5.5,6.5]~\mathrm{GeV}^{2}.
\end{equation}

\begin{figure}[h]
\includegraphics[width=8.5cm]{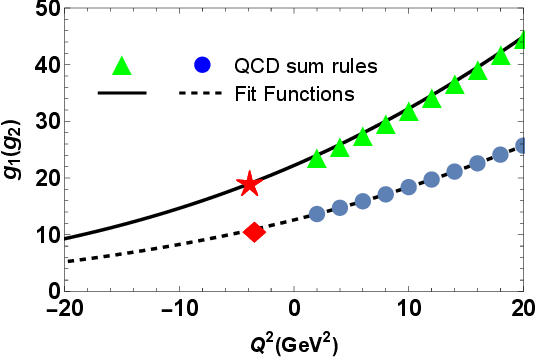}
\caption{QCD data and extrapolating functions $\mathcal{G}_1(Q^{2})$ (dashed
line) and $\mathcal{G}_2(Q^{2})$ (solid line). The diamond and star fix the
points $Q^{2}=-m_{D}^{2}$ and $Q^{2}=-m_{D_s}^{2}$ where the strong
couplings $g_1 $ and $g_2$ have been extracted. }
\label{fig:Fit}
\end{figure}

It is known that the form factor can be computed in the deep Euclidean
region $q^{2}\ll 0$. At the same time, the strong coupling $g_{1}$ required
for our purposes is defined at the mass shell of the $D^{+}$ meson, i.e., $%
g_{1}=g_{1}(m_{D}^{2})$. To escape from these problems, it is convenient to
use a variable $Q^{2}=-q^{2}$ and denote a new function as $g_{1}(Q^{2})$.
Then we calculate the form factor $g_{1}(Q^{2})$ at $Q^{2}=2-20~\mathrm{GeV}%
^{2}$ results of which are depicted in Fig.\ \ref{fig:Fit}. Afterwards, we
introduce the fit function $\mathcal{G}_{1}(Q^{2},m^{2})$ that at momenta $%
Q^{2}>0$ leads to the same SR data, but can be extrapolated to the domain of
negative $Q^{2}$. To this end, we employ the function
\begin{equation}
\mathcal{G}_{1}(Q^{2},m^{2})=\mathcal{G}_{1}^{0}\mathrm{\exp }\left[
c_{1}^{1}\frac{Q^{2}}{m^{2}}+c_{1}^{2}\left( \frac{Q^{2}}{m^{2}}\right) ^{2}%
\right] ,  \label{eq:FitF}
\end{equation}%
where $\mathcal{G}_{1}^{0}$, $c_{1}^{1}$, and $c_{1}^{2}$ are fitted
constants. Having compared the SR output and Eq.\ (\ref{eq:FitF}), it is not
difficult to find
\begin{equation}
\mathcal{G}_{1}^{0}=12.63,\text{ }c_{1}^{1}=0.69,\text{and }c_{1}^{2}=-0.06.
\label{eq:FF1}
\end{equation}%
The function $\mathcal{G}_{1}(Q^{2},m^{2})$ is also shown in Fig.\ \ref%
{fig:Fit}, where one sees a reasonable agreement with the SR data. For the
strong coupling $g_{1}$, we find
\begin{equation}
g_{1}\equiv \mathcal{G}_{1}(-m_{D}^{2},m^{2})=10.96\pm 2.32.  \label{eq:g1}
\end{equation}

The width of the decay $H_{\mathrm{c}}\rightarrow D^{+}D^{\ast }{}^{-}$ can
be obtained by means of the formula%
\begin{equation}
\Gamma \lbrack H_{\mathrm{c}}\rightarrow D^{+}D^{\ast }{}^{-}]=g_{1}^{2}%
\frac{\lambda _{1}}{40\pi ^{2}m^{2}}|M|^{2},  \label{eq:PDw2}
\end{equation}%
where $|M|^{2}$ is%
\begin{eqnarray}
&&|M|^{2}=\frac{1}{24m^{4}m_{D^{\ast }}^{2}}\left[
m^{8}-2m^{2}(2m_{D}^{2}-3m_{D^{\ast }}^{2})\right.  \notag \\
&&\times (m_{D^{\ast }}^{2}-m_{D}^{2})^{2}+(m_{D^{\ast
}}^{2}-m_{D}^{2})^{4}+m^{6}\left( 6m_{D^{\ast }}^{2}\right.  \notag \\
&&\left. \left. -4m_{D}^{2}\right) +2m^{4}(3m_{D}^{4}-8m_{D^{\ast
}}^{2}m_{D}^{2}-7m_{D^{\ast }}^{2})\right] .
\end{eqnarray}%
In Eq.\ (\ref{eq:PDw2}), we also use  the function $\lambda _{1}=\lambda
(m,m_{D^{\ast }},m_{D})$, where
\begin{eqnarray}
\lambda (a,b,c) &=&\frac{\sqrt{%
a^{4}+b^{4}+c^{4}-2(a^{2}b^{2}+a^{2}c^{2}+b^{2}c^{2})}}{2a}.  \notag \\
&&
\end{eqnarray}%
Then, for the partial width of the process under consideration, we find $\ $%
\begin{equation}
\Gamma \left[ H_{\mathrm{c}}\rightarrow D^{+}D^{\ast }{}^{-}\right]
=(41.0\pm 12.3\pm 11.5)~\mathrm{MeV}.  \label{eq:DW2}
\end{equation}%
The theoretical errors in Eq.\ (\ref{eq:DW2}) are generated by the
uncertainties in the coupling $g_{1}$, and in the masses $m$, $m_{D^{\ast }}$%
, and $m_{D}$ of the involved particles.

\subsection{$H_{\mathrm{c}}\rightarrow D_{s}^{+}D_{s}^{\ast -}$ and $%
D_{s}^{-}D_{s}^{\ast +}$}


Treatment of the decays $H_{\mathrm{c}}\rightarrow D_{s}^{+}D_{s}^{\ast -}$
and $H_{\mathrm{c}}\rightarrow D_{s}^{-}D_{s}^{\ast +}$ does not differ from
analysis presented above. Here, we concentrate on the process $H_{\mathrm{c}%
}\rightarrow D_{s}^{+}D_{s}^{\ast -}$. The correlation function necessary to
obtain the sum rule for the form factor $g_{2}(q^{2})$ is given by the
formula
\begin{eqnarray}
\Pi _{\mu \alpha \beta }(p,p^{\prime }) &=&i^{2}\int
d^{4}xd^{4}ye^{ip^{\prime }x}e^{iqy}\langle 0|\mathcal{T}\{J_{\mu
}^{D_{s}^{\ast }}(x)  \notag \\
&&\times J^{D_{s}}(y)J_{\alpha \beta }^{\dagger }(0)\}|0\rangle .
\label{eq:CF1B}
\end{eqnarray}%
In Eq.\ (\ref{eq:CF1B}) $J_{\mu }^{D_{s}^{\ast }}(x)$ and $J^{D_{s}}(x)$ are
interpolating currents for the  $D_{s}^{\ast -}$ and $D_{s}^{+}$
mesons, respectively. These currents have the following forms
\begin{equation}
J_{\mu }^{D_{s}^{\ast }}(x)=\overline{c}_{i}(x)\gamma _{\mu }s_{i}(x),\
J^{D_{s}}(x)=\overline{s}_{j}(x)i\gamma _{5}c_{j}(x).
\end{equation}%
The matrix elements to compute the phenomenological side of the sum rule are
given as
\begin{eqnarray}
0|J_{\mu }^{D_{s}^{\ast }}|D_{s}^{\ast -}(p^{\prime },\varepsilon )\rangle
&=&f_{D_{s}^{\ast }}m_{D_{s}^{\ast }}\varepsilon _{\mu }(p^{\prime }),
\notag \\
\langle 0|J^{D_{s}}|D_{s}^{+}(q)\rangle &=&\frac{f_{D_{s}}m_{D_{s}}^{2}}{%
m_{c}+m_{s}}.
\end{eqnarray}%
Here, $m_{D_{s}^{\ast }}=(2112.2\pm 0.4)~\mathrm{MeV}$, $m_{D_{s}}=(1969.0%
\pm 1.4)~\mathrm{MeV}$ and $m_{s}=(93.5\pm 0.8)~\mathrm{MeV}$ are the masses
of the mesons $D_{s}^{\ast -}$,$\ D_{s}^{+}$ and $s$ quark \cite{PDG:2024},
respectively. As the decay constants of these mesons, we employ $%
f_{D_{s}^{\ast }}=(268.8\pm 6.5)~\mathrm{MeV}$ and $f_{D_{s}}=(249.9\pm 0.5)~%
\mathrm{MeV}$.

In this case, the correlator $\Pi _{\mu \alpha \beta }^{\mathrm{Phys}%
}(p,p^{\prime })$ is given by Eq.\ (\ref{eq:PhysSide}) after evident
replacements of the masses and decay constants. The function $\Pi _{\mu
\alpha \beta }^{\mathrm{OPE}}(p,p^{\prime })$ has the form

\begin{eqnarray}
&&\Pi _{\mu \alpha \beta }^{\mathrm{OPE}}(p,p^{\prime })=-i\int
d^{4}xd^{4}ye^{ip^{\prime }x}e^{iqy}g_{s}\frac{{\lambda }_{ab}^{n}}{2}%
G_{\rho \alpha }^{n}(0)  \notag \\
&&\times \mathrm{Tr}\left[ \gamma _{\beta }S_{s}^{ij}(x-y)\gamma
_{5}S_{c}^{jb}(y)\gamma _{5}\sigma _{\mu }^{\rho }S_{c}^{ai}(-x)\right] .
\end{eqnarray}%
Then, the SR for the form factor $g_{2}(q^{2})$ is%
\begin{eqnarray}
g_{2}(q^{2}) &=&\frac{2(m_{c}+m_{s})(q^{2}-m_{D_{s}}^{2})}{\Lambda
f_{D_{s}^{\ast }}m_{D_{s}^{\ast }}f_{D_{s}}m_{D_{s}}^{2}}%
e^{m^{2}/M_{1}^{2}}e^{m_{D_{s}^{\ast }}^{2}/M_{2}^{2}}  \notag \\
&&\times \Pi _{2}(\mathbf{M}^{2},\mathbf{s}_{0}).
\end{eqnarray}%
Here
\begin{eqnarray}
&&\Pi _{2}(\mathbf{M}^{2},\mathbf{s}_{0})=\int_{4m_{c}^{2}}^{s_{0}}ds%
\int_{(m_{c}+m_{s})^{2}}^{s_{0}^{\prime }}ds^{\prime }\rho _{2}(s,s^{\prime
})  \notag \\
&&\times e^{-s/M_{1}^{2}-s^{\prime }/M_{2}^{2}}.
\end{eqnarray}%
The spectral density $\rho _{2}(s,s^{\prime })$ is composed of the
components $\rho _{2}^{\mathrm{pert.}}(s,s^{\prime })$, $\rho _{2}^{\mathrm{%
Dim3}}(s,s^{\prime })$, and $\rho _{2}^{\mathrm{Dim4}}(s,s^{\prime })$. The
term $\rho _{2}^{\mathrm{Dim3}}(s,s^{\prime })$ is proportional to the $s$%
-quark condensate $\langle \overline{s}s\rangle $. We write down below
numerical values of this and other light quark condensates that appear in
the calculations

\begin{eqnarray}
&&\langle \overline{q}q\rangle =-(0.24\pm 0.01)^{3}~\mathrm{GeV}^{3},\
\langle \overline{s}s\rangle =(0.8\pm 0.1)\langle \overline{q}q\rangle ,
\notag \\
&&\langle \overline{q}g_{s}\sigma Gq\rangle =m_{0}^{2}\langle \overline{q}%
q\rangle ,\ m_{0}^{2}=(0.8\pm 0.1)~\mathrm{GeV}^{2},  \notag \\
&&\langle \overline{s}g_{s}\sigma Gs\rangle =m_{0}^{2}\langle \overline{s}%
s\rangle .  \label{eq:Parameters}
\end{eqnarray}

In numerical analysis we use for the parameters $M_{1}^{2}$ and $s_{0}$
their values from Eq.\ (\ref{eq:Wind1}), whereas in the $D_{s}^{\ast }$
channel, we employ the regions
\begin{equation}
M_{2}^{2}\in \lbrack 2.5,3.5]~\mathrm{GeV}^{2},\ s_{0}^{\prime }\in \lbrack
6,8]~\mathrm{GeV}^{2}.
\end{equation}%
The SR data obtained for the form factor $g_{2}(Q^{2})$ at $Q^{2}=2-20~%
\mathrm{GeV}^{2}$ can be fitted by the extrapolating function $\mathcal{G}%
_{2}(Q^{2},m^{2})$ with parameters $\mathcal{G}_{2}^{0}=22.23,$ $%
c_{2}^{1}=0.68,$and $c_{2}^{2}=-0.06$. The relevant information is presented
in Fig.\ \ref{fig:Fit}.

As a result, for the strong coupling $g_{2}$, we get
\begin{equation}
g_{2}\equiv \mathcal{G}_{2}(-m_{D_{s}}^{2},m^{2})=19.0\pm 3.2.
\end{equation}%
The partial width of this process is calculated by means of Eq.\ (\ref%
{eq:PDw2}) with evident replacements of the particles' masses and parameters
$g_{1}\rightarrow g_{2}$, $\lambda _{1}\rightarrow \lambda _{2}=\lambda
(m,m_{D_{s}^{\ast }},m_{D_{s}})$. For $\Gamma \left[ H_{\mathrm{c}%
}\rightarrow D_{s}^{+}D_{s}^{\ast }{}^{-}\right] $ we find
\begin{equation}
\Gamma \left[ H_{\mathrm{c}}\rightarrow D_{s}^{+}D_{s}^{\ast }{}^{-}\right]
=(18.4\pm 4.6)~\mathrm{MeV}.  \label{eq:DW3}
\end{equation}%
The errors in Eq.\ (\ref{eq:DW3}) are total uncertainties arising from the
coupling $g_{2}$ and $m_{D_{s}^{\ast }}$, $m_{D_{s}}$.

The result Eq.\ (\ref{eq:DW3}) is also valid for the width of the second
decay $H_{\mathrm{c}}\rightarrow D_{s}^{-}D_{s}^{\ast +}$. Then, the full
width of the tensor hybrid charmonium $H_{\mathrm{c}}$ saturated by these
five processes amounts to
\begin{equation}
\Gamma \left[ H_{\mathrm{c}}\right] =(160\pm 30)~\mathrm{MeV},
\end{equation}%
which means that it is a rather broad state.


\section{Full width of $\widetilde{H}_{\mathrm{c}}$}

\label{sec:TildeHcWidth}

The hybrid charmonium $\widetilde{H}_{\mathrm{c}}$ bears the quantum numbers
$J^{\mathrm{PC}}=2^{++}$ and has the mass $\widetilde{m}=(4.5\pm 0.1)~%
\mathrm{GeV}$. These parameters enable one to fix its numerous two-meson
decay modes. It is easy to be convinced that processes $\widetilde{H}_{%
\mathrm{c}}\rightarrow D^{+}D^{-}$, $D^{0}\overline{D}^{0}$, $D^{\ast
+}D^{\ast -}$, \ $D^{\ast 0}\overline{D}^{\ast 0}$, $D_{s}^{+}D_{s}^{-}$ and
$D_{s}^{\ast +}D_{s}^{\ast -}$ are such channels.


\subsection{$\widetilde{H}_{\mathrm{c}}\rightarrow D^{+}D^{-}$ and $%
\widetilde{H}_{\mathrm{c}}\rightarrow D^{0}\overline{D}^{0}$}


To determine the width of the decay $\widetilde{H}_{\mathrm{c}}\rightarrow
D^{+}D^{-}$, we start from analysis of the correlation function
\begin{eqnarray}
\widetilde{\Pi }_{\alpha \beta }(p,p^{\prime }) &=&i^{2}\int
d^{4}xd^{4}ye^{ip^{\prime }x}e^{iqy}\langle 0|\mathcal{T}\{J^{D}(x)  \notag
\\
&&\times J^{D^{-}}(y)\widetilde{J}_{\alpha \beta }^{\dagger }(0)\}|0\rangle ,
\label{eq:CF4}
\end{eqnarray}%
with aim to find the sum rule for the form factor $\widetilde{g}_{1}(q^{2})$
and, by this way, estimate the strong coupling $\widetilde{g}_{1}$ of
particles at the vertex $\widetilde{H}_{\mathrm{c}}D^{+}D^{-}$. In Eq.\ (\ref%
{eq:CF4}), $J^{D^{-}}(x)$ is the interpolating current of the pseudoscalar $%
D^{-}$ meson
\begin{equation}
J^{D}(x)=\overline{c}_{j}(x)i\gamma _{5}d_{j}(x).
\end{equation}

The phenomenological side of this SR is given by the formula
\begin{eqnarray}
&&\widetilde{\Pi }_{\alpha \beta }^{\mathrm{Phys}}(p,p^{\prime })=\frac{%
\langle 0|J^{D}|D^{+}(p^{\prime })\rangle }{p^{\prime 2}-m_{D}^{2}}\frac{%
\langle 0|J^{D^{-}}|D^{-}(q)\rangle }{q^{2}-m_{D}^{2}}  \notag \\
&&\times \langle D^{+}(p^{\prime })D^{-}(q)|\widetilde{H}_{\mathrm{c}%
}(p,\epsilon )\rangle \frac{\langle \widetilde{H}_{\mathrm{c}}(p,\epsilon )|%
\widetilde{J}_{\alpha \beta }^{\dagger }|0\rangle }{p^{2}-\widetilde{m}^{2}}%
+\cdots .  \notag \\
&&
\end{eqnarray}%
The matrix elements of the $D^{\pm }$ mesons have simple form $%
f_{D}m_{D}^{2}/m_{c}$. The vertex $\langle D^{+}(p^{\prime })D^{-}(q)|%
\widetilde{H}_{\mathrm{c}}(p,\epsilon )\rangle $ is given by the expression
\begin{equation}
\langle D^{+}(p^{\prime })D^{-}(q)|\widetilde{H}_{\mathrm{c}}(p,\epsilon
)\rangle =\widetilde{g}_{1}(q^{2})\epsilon _{\mu \nu }(p)p^{\prime \mu
}q^{\nu }.
\end{equation}%
Then, the correlator becomes equal to
\begin{eqnarray}
&&\widetilde{\Pi }_{\alpha \beta }^{\mathrm{Phys}}(p,p^{\prime })=\frac{%
\widetilde{g}_{1}(q^{2})\widetilde{\Lambda }f_{D}^{2}m_{D}^{4}}{%
m_{c}^{2}\left( p^{2}-\widetilde{m}^{2}\right) (p^{\prime 2}-m_{D}^{2})}
\notag \\
&&\times \frac{1}{(q^{2}-m_{D}^{2})}\left\{ \frac{\widetilde{\lambda }^{2}}{3%
}g_{\alpha \beta }+\left[ \frac{m_{D}^{2}}{\widetilde{m}^{2}}+\frac{2%
\widetilde{\lambda }^{2}}{3\widetilde{m}^{2}}\right] p_{\alpha }p_{\beta
}\right.  \notag \\
&&\left. +p_{\alpha }^{\prime }p_{\beta }^{\prime }-\frac{\widetilde{m}%
^{2}+m_{D}^{2}-q^{2}}{2\widetilde{m}^{2}}(p_{\alpha }p_{\beta }^{\prime
}+p_{\beta }p_{\alpha }^{\prime })\right\} ,
\end{eqnarray}%
where $\widetilde{\lambda }=\lambda (\widetilde{m},m_{D},q)$.

In terms of the quark-gluon propagators the correlation function $\widetilde{%
\Pi }_{\alpha \beta }(p,p^{\prime })$ reads%
\begin{eqnarray}
&&\widetilde{\Pi }_{\alpha \beta }^{\mathrm{OPE}}(p,p^{\prime })=\int
d^{4}xd^{4}ye^{ip^{\prime }x}e^{iqy}g_{s}\frac{{\lambda }_{ab}^{n}}{2}%
\widetilde{G}_{\rho \alpha }^{n}(0)  \notag \\
&&\times \mathrm{Tr}\left[ \gamma _{5}S_{d}^{ij}(x-y)\gamma
_{5}S_{c}^{jb}(y)\gamma _{5}\sigma _{\beta }^{\rho }S_{c}^{ai}(-x)\right] .
\end{eqnarray}%
The SR for the form factor $\widetilde{g}_{1}(q^{2})$ is derived by
employing invariant amplitudes $\widetilde{\Pi }_{1}^{\mathrm{Phys}%
}(p^{2},p^{\prime 2},q^{2})$ and $\widetilde{\Pi }_{1}^{\mathrm{OPE}%
}(p^{2},p^{\prime 2},q^{2})$ which correspond to terms $g_{\alpha \beta }$
in the correlation functions
\begin{eqnarray}
&&\widetilde{g}_{1}(q^{2})=\frac{3m_{c}^{2}(q^{2}-m_{D}^{2})}{\widetilde{%
\Lambda }f_{D}^{2}m_{D}^{4}\widetilde{\lambda }^{2}}%
e^{m^{2}/M_{1}^{2}}e^{m_{D}^{2}/M_{2}^{2}}  \notag \\
&&\times \widetilde{\Pi }_{1}(\mathbf{M}^{2},\mathbf{s}_{0},q^{2}).
\end{eqnarray}%
Here, $\widetilde{\Pi }_{1}(\mathbf{M}^{2},\mathbf{s}_{0},q^{2})$ is the
amplitude $\widetilde{\Pi }_{1}^{\mathrm{OPE}}(p^{2},p^{\prime 2},q^{2})$
after the Borel transformations and subtractions. It is determined by the
spectral density $\widetilde{\rho }_{1}(s,s^{\prime },q^{2})$ that contains
perturbative, dimension-3, -5 and -7 terms. The dimension-5 and -7 terms are
proportional to condensates $\langle \overline{q}g_{s}\sigma Gq\rangle $ and
$\langle \alpha _{s}G^{2}/\pi \rangle \langle \overline{q}q\rangle $,
respectively.

\begin{figure}[h]
\includegraphics[width=8.5cm]{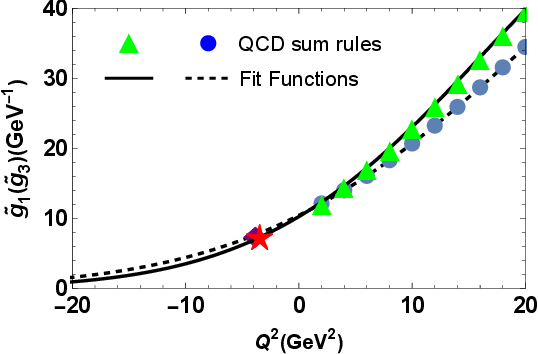}
\caption{SR data and functions $\widetilde{\mathcal{G}}_{1}(Q^{2})$ (solid
line) and $\widetilde{\mathcal{G}}_{3}(Q^{2})$ (dashed line). The labels are
placed at the points $Q^{2}=-m_{D}^{2}$ and $Q^{2}=-m_{D_{s}}^{2}$. }
\label{fig:Fit1}
\end{figure}

For the Borel and continuum subtraction parameters in the $D^{+}$ channel,
we employ%
\begin{equation}
M_{2}^{2}\in \lbrack 1.5,3]~\mathrm{GeV}^{2},\ s_{0}^{\prime }\in \lbrack
5,5.2]~\mathrm{GeV}^{2}.
\end{equation}%
The strong coupling $\widetilde{g}_{1}$ is fixed at the $D^{-}$ meson mass
shell $q^{2}=m_{D}^{2}$ by means of the extrapolating function $\widetilde{%
\mathcal{G}}_{1}(Q^{2},\widetilde{m}^{2})$ with parameters $\widetilde{%
\mathcal{G}}_{1}^{0}=10.23\ \mathrm{GeV}^{-1},$ $\widetilde{c}_{1}^{1}=1.91,$%
and $\widetilde{c}_{1}^{2}=-0.54$ (see, Fig.\ \ref{fig:Fit1}). It is worth
noting that the functions $\widetilde{\mathcal{G}}_{i}(Q^{2},\widetilde{m}%
^{2})$ are given by Eq.\ (\ref{eq:FitF}) with replacement $m\rightarrow
\widetilde{m}$.

As a result, we get
\begin{equation}
\widetilde{g}_{1}\equiv \widetilde{\mathcal{G}}_{1}(-m_{D}^{2},\widetilde{m}%
^{2})=(7.24\pm 1.41)\ \mathrm{GeV}^{-1}.
\end{equation}%
The partial width of the decay $\widetilde{H}_{\mathrm{c}}\rightarrow
D^{+}D^{-}$ is equal to
\begin{equation}
\Gamma \left[ \widetilde{H}_{\mathrm{c}}\rightarrow D^{+}D^{-}\right] =\frac{%
\widetilde{g}_{1}^{2}\widetilde{\lambda }_{1}}{960\pi \widetilde{m}^{2}}%
\left( \widetilde{m}^{2}-4m_{D}^{2}\right) ^{2},  \label{eq:PW1}
\end{equation}%
and $\widetilde{\lambda }_{1}=\lambda (\widetilde{m},m_{D},m_{D})$. Then it
is not difficult to evaluate
\begin{equation}
\Gamma \left[ \widetilde{H}_{\mathrm{c}}\rightarrow D^{+}D^{-}\right]
=(42.5\pm 16.9)~\mathrm{MeV},
\end{equation}%
where we present the total errors connected with $\widetilde{g}_{1}$ and the
masses $\widetilde{m}$,  and $m_{D}$.

The width of the process $\widetilde{H}_{\mathrm{c}}\rightarrow D^{0}%
\overline{D}^{0}$ amounts to Eq.\ (\ref{eq:PW1}) because the quark content
of the mesons $D^{0}\overline{D}^{0}$ can be obtained from $D^{+}D^{-}$
after replacement $d\rightarrow u$. Note that we ignore the small mass gap
between\ $D^{\pm }$ and $D^{0}(\overline{D}^{0})$ mesons.

\subsection{$\widetilde{H}_{\mathrm{c}}\rightarrow D^{\ast +}D^{\ast -}$, and  $%
D^{\ast 0}\overline{D}^{\ast 0}$}


Here, we consider the process $\widetilde{H}_{\mathrm{c}}\rightarrow D^{\ast
+}D^{\ast -}$ in a detailed manner. In the case of the decay $\widetilde{H}_{%
\mathrm{c}}\rightarrow D^{\ast +}D^{\ast -}$ the SR for the strong form
factor $\widetilde{g}_{2}(q^{2})$ at the vertex $\widetilde{H}_{\mathrm{c}%
}D^{\ast +}D^{\ast -}$ can be obtained from the correlation function
\begin{eqnarray}
\widetilde{\Pi }_{\mu \nu \alpha \beta }(p,p^{\prime }) &=&i^{2}\int
d^{4}xd^{4}ye^{ip^{\prime }y}e^{iqx}\langle 0|\mathcal{T}\{J_{\mu }^{D^{\ast
+}}(x)  \notag \\
&&\times J_{\nu }^{D^{\ast }}(y)\widetilde{J}_{\alpha \beta }^{\dagger
}(0)\}|0\rangle ,  \label{eq:CF7}
\end{eqnarray}%
where $J_{\mu }^{D^{\ast +}}(x)$ and $J_{\nu }^{D^{\ast }}(x)$ are the
interpolating functions of the mesons $D^{\ast }(2010)^{+}$ and $D^{\ast
}(2010)^{-}$, respectively. The current $J_{\nu }^{D^{\ast }}(x)$ has been
defined by Eq.\ (\ref{eq:CDmes}), whereas for $J_{\mu }^{D^{\ast +}}(x)$, we
have
\begin{equation}
J_{\mu }^{D^{\ast +}}(x)=\overline{d}_{i}(x)\gamma _{\mu }c_{i}(x).
\end{equation}

The correlation function $\widetilde{\Pi }_{\mu \nu \alpha \beta
}(p,p^{\prime })$ in terms of the physical parameters of the particles
involved in this decay process is
\begin{eqnarray}
&&\widetilde{\Pi }_{\mu \nu \alpha \beta }^{\mathrm{Phys}}(p,p^{\prime })=%
\frac{\langle 0|J_{\mu }^{D^{\ast +}}|D^{\ast +}(p^{\prime },\varepsilon
_{1})\rangle }{p^{\prime 2}-m_{D^{\ast }}^{2}}\frac{\langle 0|J_{\nu
}^{D^{\ast }}|D^{\ast -}(q,\varepsilon _{2})\rangle }{q^{2}-m_{D^{\ast }}^{2}%
}  \notag \\
&&\times \langle D^{\ast +}(p^{\prime },\varepsilon _{1})D^{\ast
-}(q,\varepsilon _{2})|\widetilde{H}_{\mathrm{c}}(p,\epsilon )\rangle \frac{%
\langle \widetilde{H}_{\mathrm{c}}(p,\epsilon )|\widetilde{J}_{\alpha \beta
}^{\dagger }|0\rangle }{p^{2}-\widetilde{m}^{2}}+\cdots ,  \notag \\
&&
\end{eqnarray}%
where $\varepsilon _{1\mu }$ and $\varepsilon _{2\nu }$ are the polarization
vectors of the $\ D^{\ast +}$ and $D^{\ast -}$ mesons, respectively.

The matrix elements that are used are
\begin{eqnarray}
\langle 0|J_{\mu }^{D^{\ast }+}|D^{\ast +}(p^{\prime },\varepsilon
_{1})\rangle &=&f_{D^{\ast }}m_{D^{\ast }}\varepsilon _{1\mu }(p^{\prime }),
\notag \\
\langle 0|J_{\nu }^{D^{\ast }}|D^{\ast -}(q,\varepsilon _{2})\rangle
&=&f_{D^{\ast }}m_{D^{\ast }}\varepsilon _{2\nu }(p^{\prime }),
\end{eqnarray}%
where $m_{D^{\ast }}=(1869.5\pm 0.4)\ \mathrm{MeV}$ and $f_{D^{\ast
}}=(252.2\pm 22.66)\ \mathrm{MeV}$ are the mass and decay constant of the
mesons $D^{\ast \pm }$. The vertex $\langle D^{\ast +}(p^{\prime
},\varepsilon _{1})D^{\ast -}(q,\varepsilon _{2})|\widetilde{H}_{\mathrm{c}%
}(p,\epsilon )\rangle $ has the following form
\begin{eqnarray}
&&\langle D^{\ast +}(p^{\prime },\varepsilon _{1})D^{\ast -}(q,\varepsilon
_{2})|\widetilde{H}_{\mathrm{c}}(p,\epsilon )\rangle =\widetilde{g}%
_{2}(q^{2})\epsilon _{\tau \rho }\left[ \varepsilon _{1}^{\ast }\cdot
q\right.  \notag \\
&&\left. \times \varepsilon _{2}^{\tau \ast }p^{\prime \rho }+\varepsilon
_{2}^{\ast }\cdot p^{\prime }\varepsilon _{1}^{\ast \tau }q^{\rho
}-p^{\prime }\cdot q\varepsilon _{1}^{\tau \ast }\varepsilon _{2}^{\rho \ast
}-\varepsilon _{1}^{\ast }\cdot \varepsilon _{2}^{\ast }p^{\prime \tau
}q^{\rho }\right] .  \notag \\
&&  \label{eq:TVVVertex}
\end{eqnarray}

As a result, for $\widetilde{\Pi }_{\mu \nu \alpha \beta }^{\mathrm{Phys}%
}(p,p^{\prime })$ we find the lengthy expression
\begin{eqnarray}
&&\widetilde{\Pi }_{\mu \nu \alpha \beta }^{\mathrm{Phys}}(p,p^{\prime })=%
\frac{\widetilde{g}_{2}(q^{2})\widetilde{\Lambda }f_{D^{\ast
}}^{2}m_{D^{\ast }}^{2}}{\left( p^{2}-\widetilde{m}^{2}\right) (p^{\prime
2}-m_{D^{\ast }}^{2})(q^{2}-m_{D^{\ast }}^{2})}  \notag \\
&&\times \left\{ \frac{m_{D^{\ast }}^{4}-2m_{D^{\ast }}^{2}(2\widetilde{m}%
^{2}+q^{2})+(\widetilde{m}^{2}-q^{2})(3\widetilde{m}^{2}-q^{2})}{12%
\widetilde{m}^{2}}\right.  \notag \\
&&\times g_{\mu \nu }g_{\alpha \beta }+\frac{1}{3}g_{\alpha \beta }\left(
\frac{m_{D^{\ast }}^{2}}{\widetilde{m}^{2}}p_{\mu }p_{\nu }+2p_{\mu
}^{\prime }p_{\nu }^{\prime }\right) -\frac{1}{6\widetilde{m}^{2}}g_{\alpha
\beta }  \notag \\
&&\times \left[ (m_{D^{\ast }}^{2}+3\widetilde{m}^{2}-q^{2})p_{\mu }p_{\nu
}^{\prime }+(m_{D^{\ast }}^{2}+\widetilde{m}^{2}-q^{2})p_{\mu }^{\prime
}p_{\nu }\right]  \notag \\
&&\left. +\ \text{other structures}\right\} .  \label{eq:PhysSide2}
\end{eqnarray}

For the QCD side of the sum rule, we obtain
\begin{eqnarray}
&&\widetilde{\Pi }_{\mu \nu \alpha \beta }^{\mathrm{OPE}}(p,p^{\prime
})=i^{2}\int d^{4}xd^{4}ye^{ip^{\prime }x}e^{iqy}g_{s}\frac{{\lambda }%
_{ab}^{n}}{2}\widetilde{G}_{\rho \alpha }^{n}(0)  \notag \\
&&\times \mathrm{Tr}\left[ \gamma _{\mu }S_{d}^{ij}(x-y)\gamma _{\nu
}S_{c}^{jb}(y)\gamma _{5}\sigma _{\beta }^{\rho }S_{c}^{ai}(-x)\right] .
\label{eq:CF8}
\end{eqnarray}%
The SR for the form factor $\widetilde{g}_{2}(q^{2})$ is derived using the
structures $g_{\mu \nu }g_{\alpha \beta }$ in the correlation functions.
Note that $\widetilde{g}_{2}(q^{2})$ receives contributions from the
perturbative, dimension-4, -5, and -7 terms.

In numerical analysis, the parameters $M_{2}^{2}$ and $s_{0}^{\prime }$ in
the $D^{\ast +}$ meson channel are chosen in the form
\begin{equation}
M_{2}^{2}\in \lbrack 2,4]~\mathrm{GeV}^{2},\ s_{0}^{\prime }\in \lbrack
5.5,6.5]~\mathrm{GeV}^{2}.
\end{equation}%
The strong coupling $\widetilde{g}_{2}$ amounts to
\begin{equation}
\widetilde{g}_{2}=\widetilde{\mathcal{G}}_{2}(-m_{D^{\ast }}^{2},\widetilde{m%
}^{2})=(0.84\pm 0.16)\ \mathrm{GeV}^{-1}.
\end{equation}%
It has been estimated at the mass shell $q^{2}=m_{D^{\ast }}^{2}$ of the $%
D^{\ast -}$ meson by employing the interpolating function $\widetilde{%
\mathcal{G}}_{2}(Q^{2},\widetilde{m}^{2})$. The function $\widetilde{%
\mathcal{G}}_{2}(Q^{2},\widetilde{m}^{2})$ is determined by the parameters $%
\widetilde{\mathcal{G}}_{2}^{0}=0.93\ \mathrm{GeV}^{-1},$ $\widetilde{c}%
_{2}^{1}=0.50,$ and $\widetilde{c}_{2}^{2}=-0.27$.

The width of this decay is
\begin{equation}
\Gamma \left[ \widetilde{H}_{\mathrm{c}}\rightarrow D^{\ast +}D^{\ast -}%
\right] =\frac{\widetilde{g}_{2}^{2}\widetilde{\lambda }_{2}}{80\pi
\widetilde{m}^{2}}\left( \widetilde{m}^{4}-3\widetilde{m}^{2}m_{D^{\ast
}}^{2}+6m_{D^{\ast }}^{4}\right) ,
\end{equation}%
with $\widetilde{\lambda }_{2}$ being equal to $\lambda (\widetilde{m}%
,m_{D^{\ast }},m_{D^{\ast }})$. Then, we get
\begin{equation}
\Gamma \left[ \widetilde{H}_{\mathrm{c}}\rightarrow D^{\ast +}D^{\ast -}%
\right] =(36.5\pm 11.6)~\mathrm{MeV}.
\end{equation}


\subsection{Decays to charmed-strange mesons}


The tensor hybrid state $\widetilde{H}_{\mathrm{c}}$ can also decay to
charmed-strange meson pairs $\widetilde{H}_{\mathrm{c}}\rightarrow
D_{s}^{+}D_{s}^{-}$ and $D_{s}^{\ast +}D_{s}^{\ast -}$. Similar decays have
been considered in the previous subsections. The current channels differ
from those only by parameters of the mesons $D_{s}^{\pm }$ and $D_{s}^{\ast
\pm }$. Therefore, it is enough to provide results of numerical computations.

The correlation function in the case of decay $\widetilde{H}_{\mathrm{c}%
}\rightarrow D_{s}^{+}D_{s}^{-}$ contains contributions of the perturbative
and dimension-3, -4, -5 and-7 terms which are proportional to $\langle
\overline{s}s\rangle $, $\langle \overline{s}g_{s}\sigma Gs\rangle $, $%
\langle \alpha _{s}G^{2}/\pi \rangle $ and $\langle \alpha _{s}G^{2}/\pi
\rangle \langle \overline{s}s\rangle $ condensates, respectively. The
coupling $\widetilde{g}_{3}$ responsible for strong interactions of
particles at the vertex $\widetilde{H}_{\mathrm{c}}D_{s}^{+}D_{s}^{-}$ is
\begin{equation}
\widetilde{g}_{3}\equiv \widetilde{\mathcal{G}}_{3}(-m_{D_{s}}^{2},%
\widetilde{m}^{2})=(7.60\pm 1.28)\ \mathrm{GeV}^{-1}.
\end{equation}%
The fitting function $\widetilde{\mathcal{G}}_{3}(Q^{2})$ is determined by
the coefficients $\widetilde{\mathcal{G}}_{3}^{0}=10.41\ \mathrm{GeV}^{-1},$
$\widetilde{c}_{3}^{1}=1.57,$and $\widetilde{c}_{2}^{2}=-0.36$. The relevant
SR predictions  for the form factor $\widetilde{g}_{3}(Q^{2})$ and the function $\widetilde{\mathcal{G}}_{3}(Q^{2})$ are also plotted in Fig.\ \ref{fig:Fit1}%
. The width of the decay $\widetilde{H}_{\mathrm{c}}\rightarrow
D_{s}^{+}D_{s}^{-}$ is

\begin{equation}
\Gamma \left[ \widetilde{H}_{\mathrm{c}}\rightarrow D_{s}^{+}D_{s}^{-}\right]
=(23.3\pm 11.4)~\mathrm{MeV}.
\end{equation}

The process $\widetilde{H}_{\mathrm{c}}\rightarrow D_{s}^{\ast +}D_{s}^{\ast
-}$ is determined by the following parameters:%
\begin{equation}
\widetilde{g}_{4}\equiv \widetilde{\mathcal{G}}_{4}(-m_{D_{s}^{\ast }}^{2},%
\widetilde{m}^{2})=(0.78\pm 0.13)\ \mathrm{GeV}^{-1},
\end{equation}%
and

\begin{equation}
\Gamma \left[ \widetilde{H}_{\mathrm{c}}\rightarrow D_{s}^{\ast
+}D_{s}^{\ast -}\right] =(24.1\pm 11.5)~\mathrm{MeV}.
\end{equation}

Decays considered in the present section allow us to estimate the full decay
width of the tensor hybrid charmonium $\widetilde{H}_{\mathrm{c}}$, which
amounts to
\begin{equation}
\Gamma \left[ \widetilde{H}_{\mathrm{c}}\right] =(206\pm 33)~\mathrm{MeV}.
\end{equation}


\section{Discussion and concluding notes}

\label{sec:Dis} 
In the present article, we have evaluated the parameters of the tensor
hybrid charmonia $H_{\mathrm{c}}$ and $\widetilde{H}_{\mathrm{c}}$. Our
predictions for the masses of these states $m=(4.16\pm 0.14)~\mathrm{GeV}$
and $\widetilde{m}=(4.5\pm 0.1)~\mathrm{GeV}$ demonstrate that these
structures are unstable against strong decays to conventional meson pairs.

The masses of the tensor hybrid charmonia $H_{\mathrm{c}}$ and $\widetilde{H}%
_{\mathrm{c}}$ were computed in numerous publications. Results some of these
works are collected in Table\ \ref{tab:Compar}. In the framework of the sum
rule method these particles were investigated in Refs.\ \cite%
{Chen:2013zia,Wang:2024hvp}. A comparison of our results for $m$ and $%
\widetilde{m}$ and those of Ref.\ \cite{Chen:2013zia} reveals that $m$ is
larger than prediction made there, whereas $\widetilde{m}$ is in nice
agreement with the results of the same paper. The masses extracted in Ref.\
\cite{Wang:2024hvp} exceed our results for $m$ and $\widetilde{m}$.

\begin{table}[tbp]
\begin{tabular}{|c|c|c|}
\hline\hline
Publications & $J^{\mathrm{PC}}=2^{-+}$ & $J^{\mathrm{PC}}=2^{++}$ \\ \hline
This work & $4.16 \pm 0.14$ & $4.5 \pm 0.1$ \\
Ref.\ \cite{Chen:2013zia} & $4.04 \pm 0.23$ & $4.45 \pm 0.27$ \\
Ref.\ \cite{Wang:2024hvp} & $4.31\pm ~0.08$ & $4.85\pm ~0.06$ \\
Ref.\ \cite{HadronSpectrum:2012gic} & $4.334 \pm ~0.017$ & $4.492\pm ~0.021$
\\
Ref.\ \cite{Cheung:2016bym} & $4.456 \pm ~0.021$ & $4.623\pm ~0.032$ \\
Ref.\ \cite{Berwein:2015vca} $V^{(0.5)}$ & $4.05 \pm 0.15$ & $4.17 \pm 0.15$
\\
Ref.\ \cite{Berwein:2015vca} $V^{(0.25)}$ & $4.15 \pm 0.15$ & $4.37 \pm 0.15$
\\
Ref.\ \cite{Soto:2023lbh} & $4.046 \pm 0.030$ & $4.232\pm 0.030$ \\
\hline\hline
\end{tabular}%
\caption{Predictions for the masses of the tensor $\overline{c}gc$ hybrids $%
J^{\mathrm{PC}}=2^{-+}$ and $J^{\mathrm{PC}}=2^{++}$. All masses are given
in units of $\mathrm{GeV}$. }
\label{tab:Compar}
\end{table}

The hybrid mesons $H_{\mathrm{c}}$ and $\widetilde{H}_{\mathrm{c}}$ were
examined in the context of the alternative methods as well. In the context
of the lattice simulations these tensor particles have approximately the
masses $4.33~\mathrm{GeV}$ and $4.49~\mathrm{GeV}$ \cite%
{HadronSpectrum:2012gic}, and $4.46~\mathrm{GeV}$ and $4.62~\mathrm{GeV}$
\cite{Cheung:2016bym}, respectively. The lattice calculations lead, as
usual, to larger predictions: Only the hybrid $\widetilde{H}_{\mathrm{c}}$
from Ref.\ \cite{HadronSpectrum:2012gic} has a mass compatible with $%
\widetilde{m}$.

The nonrelativistic effective field theory elaborated in Ref.\ \cite%
{Berwein:2015vca} generates two sets of the masses for the charmonium
hybrids with diverse spin-parities. These sets depend on the choice of the
potential $V^{(0.5)}$ or $V^{(0.25)}$ in the coupled Schrodinger equation.
Predictions for the masses of the hybrids $H_{\mathrm{c}}$ and $\widetilde{H}%
_{\mathrm{c}}$ are shown in Table\ \ref{tab:Compar}. It is evident that
within the theoretical errors results of the potential $V^{(0.25)}$ are
consistent with our findings. Within the BOEFT method the charmonium hybrids
were examined in Ref.\ \cite{Soto:2023lbh}. Our results $m$ and $\widetilde{m%
}$ exceed the mass of the tensor $\overline{c}gc$ hybrids presented there.
Evidently, there is a necessity to continue studies to improve agreement
between diverse predictions.

Another important measurable parameter of the hybrid mesons is the decay
widths of these structures. Information on decay pattern and full width of
the hybrid states may allow one to distinguish them from other exotic
mesons. For  a long time it was believed that hybrid states were forbidden
to decay into two $S$-wave mesons \cite{Page:1996rj}. Such conclusion was
made in the context of the constituent gluon and flux tube models. Recently,  an
analysis based on the BO approximation demonstrated that decays into two $S$%
-wave heavy-light mesons are allowed for quarkonium hybrids for a wide range
of the spin-parities $J^{\mathrm{PC}}$, including tensor particles with ${%
2^{-+}}$ and ${2^{++}}$ \cite{Bruschini:2023tmm}.
nnumerical
In the present paper decays of the tensor charmonia have been examined
within the QCD sum rule framework. Note that the QCD sum rule method relies on
first principles of the field theory, and treats the heavy and light quarks
as well as the gluon field on the same footing. All nonperturbative
information required to calculate parameters of a decay is encoded in the
vacuum expectation values of numerous quark, gluon and mixed operators.
These condensates are universal parameters the  numerical values of which were
extracted from analyses of different processes.

In our article, for the first time, we calculated widths $\Gamma \left[
\widetilde{H}_{\mathrm{c}}\right] $ and $\Gamma \left[ H_{\mathrm{c}}\right]
$ of the tensor hybrids in the context of the SR method. In the case of the
structure $H_{\mathrm{c}}$ we have studied the decays $H_{\mathrm{c}%
}\rightarrow D^{(\pm )}D^{\ast }{}^{(\mp )}$, $D^{0}\overline{D}^{\ast
}{}^{0}$, and $D_{s}^{(\pm )}D_{s}^{\ast }{}^{(\mp )}$. The full width of $%
H_{\mathrm{c}}$ saturated by these decay channels amounts to $\Gamma \left[
H_{\mathrm{c}}\right] =(160\pm 30)~\mathrm{MeV}$. The width of the hybrid
state $\widetilde{H}_{\mathrm{c}}$ is equal to $\Gamma \left[ \widetilde{H}_{%
\mathrm{c}}\right] =(206\pm 33)~\mathrm{MeV}$ which have been evaluated by
taking into account the kinematically allowed processes $\widetilde{H}_{%
\mathrm{c}}\rightarrow D^{+}D^{-}$, $D^{0}\overline{D}^{0}$, $D^{\ast
+}D^{\ast -}$, $D^{\ast 0}\overline{D}^{\ast 0}$, $D_{s}^{+}D_{s}^{-}$ and $%
D_{s}^{\ast +}D_{s}^{\ast -}$. Clearly, both of these hypothetical hybrid
charmonia are broad structures. Predictions for these parameters can be
refined by taking into consideration their other decay modes.

The diverse decay channels of the heavy hybrid mesons were previously
examined in Refs.\ \cite%
{Brambilla:2022hhi,TarrusCastella:2021pld,Braaten:2024stn,TarrusCastella:2024zps}%
, as well. In Refs. \cite{Brambilla:2022hhi,TarrusCastella:2021pld}
the authors concentrated on transitions of the heavy hybrids to standard
quarkonia. The most general consideration of these decays was done in Ref.\
\cite{Brambilla:2022hhi}, in which BOEFT was applied to investigate the
semi-inclusive transition rate of a heavy quarkonium hybrid $H_{\mathrm{m}}$
to $Q_{\mathrm{n}}+X$ with $Q_{\mathrm{n}}$ and $X$ being a quarkonium and
any final state of light particles, respectively. It turns out that the
widths of such transitions are sizeable. For instance, width of the decay $%
H_{\mathrm{c}}\rightarrow \eta _{c}(1S)+X$ is $35_{-9-12}^{+15+16}~\mathrm{%
MeV}$, whereas for $\widetilde{H}_{\mathrm{c}}\rightarrow \chi _{c}(1P)+X$ 
the authors obtained $65_{18-28}^{+41+40}~\mathrm{MeV}$.

The selection rules and relative partial decay rates for decays of
quarkonium hybrids with various $J^{\mathrm{PC}}$ into pairs of heavy-light
mesons were investigated in Ref.\ \cite{Braaten:2024stn}. It is worth noting
that analysis was performed by employing the Born-Oppenheimer approximation.
Decays of hybrids were treated through transitions from confining BO
potentials to meson-pair potentials with the same BO quantum numbers.
Numerical computations were carried out in the case of the decays into pairs
of $B^{(\ast)}$ mesons. We assume that the same ratios are valid for
decays of the hidden-charm hybrids into $D$ meson pairs. Then, in the case
of the hybrid ${2^{++}}$, Ref.\ \cite{Braaten:2024stn} predicted for decays
(in $S+D$ wave) into mesons $D\overline{D}$, $D\overline{D}^{\ast }+%
\overline{D}D^{\ast }$ and $\overline{D}^{\ast }D^{\ast }$ the ratios $3:9:12
$, i.e., decays to vector mesons are dominant channels for $\widetilde{H}_{%
\mathrm{c}}$ state. Our results imply fulfilment of the approximate ratios $%
3.5:3:1:1$ for decays through channels $D^{+}D^{-}+D^{0}\overline{D}^{0}$, $%
D^{\ast +}D^{\ast -}+D^{\ast 0}\overline{D}^{\ast 0}$, $D_{s}^{+}D_{s}^{-}$
and $D_{s}^{\ast +}D_{s}^{\ast -}$, respectively. As is seen, in the
framework of QCD sum rule approach decays of the tensor hybrid charmonia $%
\widetilde{H}_{\mathrm{c}}$ into pseudoscalar and vector meson pairs
contribute equally to its full width.

The investigation carried out in this paper has demonstrated that there is
necessity to continue analyses of the heavy hybrid mesons in the context of
different method to refine and correct predictions for parameters of these
structures. This is required also to remove observed discrepancies in outputs
of various methods.

\section*{ACKNOWLEDGMENTS}

K. Azizi is grateful to the Iran National Science Foundation (INSF) for the
partial financial support provided under the Elites Grant No. 4037888.

\end{document}